\documentclass[12pt]{article}
\pdfoutput=1

\usepackage{amsmath}
\usepackage{amssymb}
\usepackage{epsfig}
\usepackage{epstopdf}
\usepackage{latexsym}
\usepackage{color}
\usepackage{slashed}
\usepackage{tensor}
\numberwithin{equation}{section}
\usepackage[
      colorlinks=false,
      linkcolor=darkblue,  
      urlcolor=blue,    
      filecolor=blue,     
      citecolor=red,
linktocpage=true,
      pdfstartview=FitV,
      bookmarksopen=true    
      ]{hyperref}

\newcommand{\dd}{d}
\newcommand{\be}{\begin{equation}}
\newcommand{\ee}{\end{equation}}
\newcommand{\ii}{\mathrm{i}}
\newcommand{\me}{\mathrm{e}}
\usepackage{multirow}
\usepackage[bbgreekl]{mathbbol}
\usepackage{bbold}
\DeclareSymbolFontAlphabet{\mathbbl}{bbold}
\begin{document}
%%%%%%%%%%%%%%%%%%%%%%%%%%%%%%%%%%%%%

%%%%%%%%%%%%%%%%%%%%%%%%%%%%%%%%%%%%%
\begin{titlepage}
%%%%%%%%%%%%%%%%%%%%%%%%%%%%%%%%%%%%%

\centerline{}
\centerline{}

\centerline{\Huge \rm D4-branes wrapped on four-dimensional orbifolds} 
\bigskip
\centerline{\Huge \rm through consistent truncation} 

\vspace{18mm}
\centerline{Christopher Couzens$^{\text{a}}$, Hyojoong Kim$^{\text{b}}$, Nakwoo Kim$^{\text{b}}$, Yein Lee$^{\text{b}}$ and Minwoo Suh$^{\text{b}}$}

\vspace{8mm}

\centerline{$^\text{a}$\it{Mathematical Institute, University of Oxford}}
\medskip

\centerline{\it{ Andrew Wiles Building, Radcliffe Observatory Quarter}}

\medskip
\centerline{\it{Woodstock Road, Oxford, OX2 6GG, U.K.}}

\vspace{4mm}
\centerline{ $^{\text{b}}$\it{Department of Physics and Research Institute of Basic Science}}
\medskip
\centerline{\it {Kyung Hee University, Seoul 02447, Korea}}

\vspace{5mm}

\centerline{\tt Christopher.Couzens@maths.ox.ac.uk, h.kim@khu.ac.kr, nkim@khu.ac.kr} 
\centerline{\tt lyi126@khu.ac.kr, minwoosuh1@gmail.com} 

\vspace{12mm}

\begin{abstract}
\noindent We construct a consistent truncation of six-dimensional matter coupled $F(4)$ gauged supergravity on a cornucopia of two-dimensional surfaces including a spindle, disc, domain wall and other novel backgrounds to four-dimensional minimal gauged supergravity. Using our consistent truncation we uplift known AdS$_2\times \mathbbl{\Sigma}_1$ solutions giving rise to four-dimensional orbifold solutions, AdS$_2\times\mathbbl{\Sigma}_1\ltimes\mathbbl{\Sigma}_2$. We further uplift our solutions to massive type IIA supergravity by constructing the full uplift formulae for six-dimensional U$(1)^2$-gauged supergravity including all fields and arbitrary Romans mass and gauge coupling. The solutions we construct are naturally interpreted as the near-horizon geometries of asymptotically AdS$_6$ black holes with a four-dimensional orbifold horizon. Alternatively, one may view them as the holographic duals of superconformal quantum mechanical theories constructed by compactifying five-dimensional USp$(2N)$ theory living on a stack of D4-D8 branes on the four-dimensional orbifolds. As a first step to identifying these quantum mechanical theories we compute the Bekenstein--Hawking entropy holographically. 

\end{abstract}

\vskip 2cm

%t\flushleft {October, 2022}

%%%%%%%%%%%%%%%%%%%%%%%%%%%%%%%%%%%%%
\end{titlepage}
%%%%%%%%%%%%%%%%%%%%%%%%%%%%%%%%%%%%%

\tableofcontents

%%%%%%%%%%%%%%%%%%%%%%%%%%%%%%%%%%%%%
\section{Introduction}
%%%%%%%%%%%%%%%%%%%%%%%%%%%%%%%%%%%%%

Compactifying a field theory using a topological twist, \cite{Witten:1988ze}, has long been a useful tool for studying supersymmetric field theories and their holographic duals, \cite{Maldacena:2000mw}. Recently, a novel class of anti-de Sitter solutions were discovered which realise supersymmetry using a different type of twisting. The solutions are obtained from wrapping branes on the orbifold, $\mathbbl{\Sigma}\equiv\mathbbl{WCP}^{1}_{[n_-,n_+]}$, known as a spindle. The spindle has conical deficit angles at both poles with weights, $n_{\pm}$, with $n_{\pm}$ relatively prime integers. The first spindle solutions in the supergravity literature describe D3-branes wrapped on a spindle, \cite{Ferrero:2020laf}. Since then, spindle solutions have been extended to many other brane configurations and multi-charged solutions: D3-branes, \cite{Hosseini:2021fge, Boido:2021szx}, M2-branes, \cite{Ferrero:2020twa, Cassani:2021dwa, Ferrero:2021ovq, Couzens:2021rlk,Faedo:2021kur}, M5-branes, \cite{Ferrero:2021wvk}, D4-branes, \cite{Faedo:2021nub, Giri:2021xta}, mass-deformed D3-branes, \cite{Arav:2022lzo}, D2-branes, \cite{Couzens:2022yiv}, and more general massive type IIA brane configurations in \cite{Cheung:2022wpg, Couzens:2022aki}. All of these solutions exhibit a universal feature. Supersymmetry is preserved by one of two new types of twist, dubbed \emph{twist} or \emph{anti-twist}. 
In \cite{Ferrero:2021etw} it was shown that these are the only two possible twists on a spindle that preserve supersymmetry.\footnote{In \cite{Couzens:2021cpk} infinite families of examples of M2-brane solutions admitting both twist and anti-twist solutions were constructed, complementing the analysis in \cite{Ferrero:2021etw}. It is interesting to note that of the known solutions only the M2-brane and D3-brane solutions allow for both twists, this is not to say that the other configurations cannot allow both though. }

In a somewhat parallel development, orbifold solutions known as discs were constructed in \cite{Bah:2021mzw, Bah:2021hei} describing M5-branes wrapped on a punctured sphere. These AdS$_5$ solutions are conjectured to be dual to a class of 4d $\mathcal{N}=2$ Argyres-Douglas theories, \cite{Argyres:1995jj}, with further checks of the dualities and generalisations of the punctures performed in \cite{Couzens:2022yjl, Bah:2022yjf}. Like the spindles this has been extended to other brane configurations: D3-branes, \cite{Couzens:2021tnv, Suh:2021ifj}, M2-branes, \cite{Suh:2021hef, Couzens:2021rlk}, D4-branes, \cite{Suh:2021aik}, and M5-branes, \cite{Karndumri:2022wpu}. In \cite{Couzens:2021tnv,Couzens:2021rlk} it was shown that the disc solutions are different global completions of the same local solution. This fact will appear later in this work where we extend it to additional types of spaces including the defect solutions of \cite{Gutperle:2022pgw}.

A natural generalisation of this program is to find AdS solutions obtained from branes wrapped on higher-dimensional orbifolds. A simple higher-dimensional oribfold is given by the product of a spindle and a constant curvature manifold. In this spirit, AdS$_2\times\mathbbl{\Sigma}\times\Sigma_{\mathfrak{g}}$ solutions in six-dimensional gauged supergravity were constructed in \cite{Giri:2021xta, Faedo:2021nub,Suh:2022olh}, while in seven-dimensional gauged supergravity, AdS$_3\times\mathbbl{\Sigma}\times\Sigma_{\mathfrak{g}}$ solutions with $\Sigma_{\mathfrak{g}}$ a Riemann surface of genus, $\mathfrak{g}$, were constructed in \cite{Suh:2022olh,Boido:2021szx} and AdS$_2\times\mathbbl{\Sigma}\times\mathbb{H}^3$ solutions in \cite{Couzens:2022yiv}. These solutions can be viewed as orbifold generalisations of the solutions obtained from branes wrapped on a product of constant curvature manifolds: AdS$_2\times\Sigma_{\mathfrak{g}_1}\times\Sigma_{\mathfrak{g}_2}$ solutions in six-dimensional gauged supergravity, \cite{Suh:2018tul, Hosseini:2018usu, Suh:2018szn}, and AdS$_3\times\Sigma_{\mathfrak{g}_1}\times\Sigma_{\mathfrak{g}_2}$ and AdS$_2\times\Sigma_{\mathfrak{g}}\times\mathbb{H}^3$ solutions in seven-dimensional gauged supergravity, \cite{Gauntlett:2001jj}.

More recently, this has been further extended in \cite{Cheung:2022ilc} to AdS$_3\times\mathbbl{\Sigma}_1\ltimes\mathbbl{\Sigma}_2$ and AdS$_3\times\Sigma_{\mathfrak{g}}\ltimes\mathbbl{\Sigma}_2$ solutions in seven-dimensional gauged supergravity where the two factors are fibred over each other. These solutions were found by performing a consistent truncation on the AdS$_5\times\mathbbl{\Sigma}$ solutions in seven-dimensional U$(1)^2$-gauged supergravity of \cite{Ferrero:2021wvk} to five-dimensional minimal gauged supergravity. Utilising the truncation ansatz the authors uplifted solutions of five-dimensional minimal gauged supergravity to seven dimensions leading to the aforementioned fibered solutions. 

In this paper we will perform the analogous construction of \cite{Cheung:2022ilc} for six-dimensional U$(1)^2$-gauged supergravity. We construct a consistent truncation of six-dimensional U$(1)^2$-gauged supergravity down to four-dimensional minimal gauged supergravity by compactifying on a local solution which contains the spindle as a choice of global completion. We study two classes of the multitude of possible solutions, focussing on AdS$_2\times\mathbbl{\Sigma}_1\ltimes\mathbbl{\Sigma}_2$ and AdS$_2\times\Sigma_{\mathfrak{g}}\ltimes\mathbbl{\Sigma}_2$ solutions. We further uplift the solutions to massive type IIA supergravity on the Brandhuber--Oz solution, \cite{Brandhuber:1999np}. For the vanishing two-form field, $B\,=\,0$, the uplift formula of six-dimensional U$(1)^2$-gauged supergravity to massive type IIA supergravity was first presented in \cite{Cvetic:1999xx} and improved in \cite{Faedo:2021nub}, but the full truncation was lacking. We have remedied this, constructing the most general uplift of the U$(1)^2$ theory.

The layout of this paper is as follows. In section \ref{sec:truncation} we perform a consistent KK reduction of six-dimensional U$(1)^2$-gauged supergravity to four-dimensional minimal gauged supergravity. The choice of two-dimensional surface on which we compactify is large, including, but not restricted to, spindles and discs. In section \ref{sec:spinspin} we use our consistent truncation to construct AdS$_2\times\mathbbl{\Sigma}_1\ltimes\mathbbl{\Sigma}_2\ltimes\hat{S}^4$ solutions of massive type IIA supergravity, with $\hat{S}^4$ a four-dimensional hemisphere. We analyse the global properties of the solutions before computing their Bekenstein--Hawking entropy, giving a prediction to match with a yet to be found dual field theory. In section \ref{sec:spinriem}, using our truncation we construct AdS$_2\times\Sigma_{\mathfrak{g}}\ltimes\mathbbl{\Sigma}_2$ solutions. We conclude in section \ref{sec:conclude}. Some technical details are relegated to five appendices. In appendix \ref{app:EOMs} we present the conventions of six-dimensional U$(1)^2$-gauged supergravity that we use in the main text. In appendix \ref{app:Uplift} we present the fully general uplift of six-dimensional U$(1)^2$-gauged supergravity to massive type IIA on the Brandhuber--Oz solution.  Appendix \ref{app:globalsols} studies the zoo of all solutions on which we may perform our truncation. Some of the solutions presented there have not appeared in the literature previously and are deserving of further study independent of these results. Appendix \ref{app:susy} proves that our consistent truncation preserves supersymmetry provided superysmmetry is preserved in four dimensions. Finally, in appendix \ref{app:Rsymmetry} we study the R-symmetry of the four-dimensional orbifolds that we construct.

\bigskip

\noindent {\bf Note added:} While we were in a process of submitting we became aware of the work of \cite{Faedo:2022aaa} which has a large overlap with some of the results presented here. For this reason, we coordinated the submission to arXiv of our respective papers.

%%%%%%%%%%%%%%%%%%%%%%%%%%%%%%%%%%%%%
\section{Consistent truncation on a two-dimensional surface}\label{sec:truncation}
%%%%%%%%%%%%%%%%%%%%%%%%%%%%%%%%%%%%%

In this section, we perform a consistent truncation of matter coupled $F(4)$ gauged supergravity in six dimensions on a two-dimensional surface down to four-dimensional minimal gauged $\mathcal{N}=2$ supergravity. The seed solutions on which we truncate down to four dimensions are the local AdS$_4\times M_2$ solutions studied in \cite{Faedo:2021nub}, of which a particular global completion is the spindle, $M_2=\mathbbl{\Sigma}_2$. We begin this section by fixing our conventions of matter coupled $F(4)$ gauged supergravity, then study the local solutions on which we will truncate to four-dimensional minimal gauged supergravity. There are a plethora of different global completions, including but not restricted to, spindles, discs, domain walls and constant curvature Riemann surfaces. Much of the global analysis of the zoo of solutions is relegated to appendix \ref{app:globalsols}, leaving just the analysis of the spindle in this section, since it is the space we will use when constructing explicit examples in the later sections. Having discussed the local and global form of the two-dimensional surface, we construct a consistent truncation of the six-dimensional theory down to four-dimensional minimal gauged $\mathcal{N}=2$ supergravity. Proof of supersymmetry preservation of the truncation is also relegated to an appendix, in this case, appendix \ref{app:susy}.

%%%%%%%%%%%%%%%%%%%%%%%%%%%%%%%%%%%%%
\subsection{U$(1)^2$-gauged supergravity in six dimensions}
%%%%%%%%%%%%%%%%%%%%%%%%%%%%%%%%%%%%%

We will consider the truncation of $F(4)$ gauged supergravity, \cite{Romans:1985tw}, coupled to matter multiplets, \cite{Andrianopoli:2001rs}, down to four-dimensional minimal gauged supergravity. More concretely, we consider pure six-dimensional SU$(2)\times{\text{U}}(1)$-gauged supergravity coupled to a vector multiplet and then truncate to U$(1)^2$-gauged supergravity by gauging the Cartan of the SU$(2)$ gauge group and a U$(1)$ from the vector multiplet, \cite{Karndumri:2015eta}. This truncation was employed to construct black hole solutions with a horizon of the form AdS$_2\times\Sigma_{\mathfrak{g}_1}\times\Sigma_{\mathfrak{g}_2}$ in \cite{Hosseini:2018usu, Suh:2018szn}.\footnote{Note that this includes AdS$_2\times M_4$ solutions, with $M_4$ K\"ahler--Einstein, as a subclass.} We follow the conventions of \cite{Faedo:2021nub} for the supergravity theory. 

The bosonic field content of the truncated theory consists of the metric, two U$(1)$ gauge fields, $A_i$, a two-form field, $B$, and two scalar fields, $\varphi_i$, with $i\,=\,1,2$. It is useful to parametrise the scalar fields by
\begin{equation}
X_i\,=\,e^{-\frac{1}{2}\vec{a}_i\cdot\vec{\varphi}}\,, \qquad \vec{a}_1\,=\,\left(2^{1/2},2^{-1/2}\right)\,, \qquad \vec{a}_2\,=\,\left(-2^{1/2},2^{-1/2}\right)\,,
\end{equation}
and to define
\begin{equation}
X_0\,=\,\frac{3m}{2g}\left(X_1X_2\right)^{-3/2}\,.
\end{equation}
The field strengths of the gauge fields and two-form field are, respectively,
\begin{equation}
F_i\,=\,dA_i\,, \qquad H=dB\,.
\end{equation}
The action is
\begin{align} \label{actionaction}
S\,&=\,\frac{1}{16\pi{G}_N^{(6)}}\int{d}^6x\sqrt{-g}\left[R-V-\frac{1}{2}|d\vec{\varphi}|^2-\frac{1}{2}\sum_{i=1}^2X_i^{-2}|F_i|^2-\frac{1}{8}\left(X_1X_2\right)^2|H|^2\right. \notag \\
&\left.-\frac{m^2}{4}\left(X_1X_2\right)^{-1}|B|^2-\frac{1}{16}\frac{\varepsilon^{\mu\nu\rho\sigma\tau\lambda}}{\sqrt{-g}}B_{\mu\nu}\left(F_{1\rho\sigma}F_{2\tau\lambda}+\frac{m^2}{12}B_{\rho\sigma}B_{\tau\lambda}\right)\right]\,,
\end{align}
where the scalar potential is
\begin{equation}
g^{-2}V\,=\,\frac{4}{9}X_0^2-4X_1X_2-\frac{8}{3}X_0\left(X_1+X_2\right)\,.
\end{equation}
For the norm of the form fields appearing in the action we use the canonical convention to include the numerical weighted factor, $i.e.$,
\begin{equation}
|\omega|^2\,=\,\frac{1}{p!}\omega_{\mu_1\ldots\mu_p}\omega^{\mu_1\ldots\mu_p}\,.
\end{equation}
The equations of motion following from the action are presented in appendix \ref{app:EOMs}. 

As we have in mind the uplift of supersymmetric AdS vacua to ten dimensions, we only consider the case of $m,g>0$ in this paper. There are two critical points of the scalar potential $V$: $X_1=X_2=(3m/2g)^{1/4}$ which is supersymmetric and $X_1=X_2=(m/2g)^{1/4}$ which is non-supersymmetric, \cite{Romans:1985tw}, and unstable, \cite{Suh:2020rma}. One typically sets the scalar fields to have value $1$ at the supersymmetric fixed point which fixes $3m=2g$.
The uplift formula to massive type IIA that we present in appendix \ref{app:Uplift} does not fix a relation between $m$ and $g$ and therefore can be used to uplift both the supersymmetric and non-supersymmetric solutions to ten dimensions. One must be slightly careful when either $g$ or $m$ vanishes or become negative, however, by using a scaling symmetry of the solution that we give explicitly, one can also uplift for these choices.

%%%%%%%%%%%%%%%%%%%%%%%%%%%%%%%%%%%%%
\subsection{AdS$_4\times M_2$ solutions}\label{sec:M2sols}
%%%%%%%%%%%%%%%%%%%%%%%%%%%%%%%%%%%%%

We now discuss the seed AdS$_4\times M_2$ solution on which we may construct the consistent truncation. Though our focus will be on obtaining a consistent truncation on a spindle, since the results are independent of the global data of $M_2$, only depending on the local form of the metric, gauge fields and scalar fields, it will be equally valid for any other global completion of the local solution. In the following we will present the local solution before studying the various ways of globally completing the solution. We relegate technical analysis of the various global solutions to appendix \ref{app:globalsols}. As we will show, there is a whole zoo of globally well-defined solutions of differing topologies that arise from the same local solution, some of which have not been studied previously. 

The local solution we consider is obtained by a double analytic continuation of a black hole solution in \cite{Cvetic:1999un}. In  \cite{Faedo:2021nub} it was shown that one choice of global completion of the solution is a spindle. We will follow their conventions in the following. The metric, the gauge fields and the scalar fields are, respectively,
\begin{align} \label{ads4sigma2}
ds_6^2\,=&\,\left(y^2h_1(y)h_2(y)\right)^{1/4}\left[ds_{\text{AdS}_4}^2+ds_{{M}_2}^2\right]\,, \notag \\
A_i\,=&\,\left(\alpha_i-\frac{y^3}{h_i(y)}\right)dz\,, \qquad X_i\,=\,\left(y^2h_1(y)h_2(y)\right)^{3/8}h_i(y)^{-1}\,,
\end{align}
where the metric on the two-dimensional surface is given by
\begin{equation}
ds_{M_2}^2\,=\,\frac{y^2}{F(y)}dy^2+\frac{F(y)}{h_1(y)h_2(y)}dz^2\,.
\end{equation}
The constants, $\alpha_i$, are pure gauge and have been introduced for later exposition. The two-form field, $B$, is trivial for the solution. This is a supersymmetric solution of six-dimensional U$(1)^2$-gauged supergravity reviewed in the previous section provided that the functions are
\begin{align}
F(y)\,=&\,m^2h_1(y)h_2(y)-y^4\,, \notag \\
h_i(y)\,=&\,\frac{2g}{3m}y^3+q_i\,, 
\end{align}
where $q_i$, $i=1,2$, are real parameters. In the following we will set $3m=2g$ periodically. This is the value which allows for the supersymmetric AdS$_6$ vacuum and is thus the necessary one to uplift if we want to preserve supersymmetry in ten dimensions.

\begin{figure}[h!]
\centering
\includegraphics[width=0.7\linewidth]{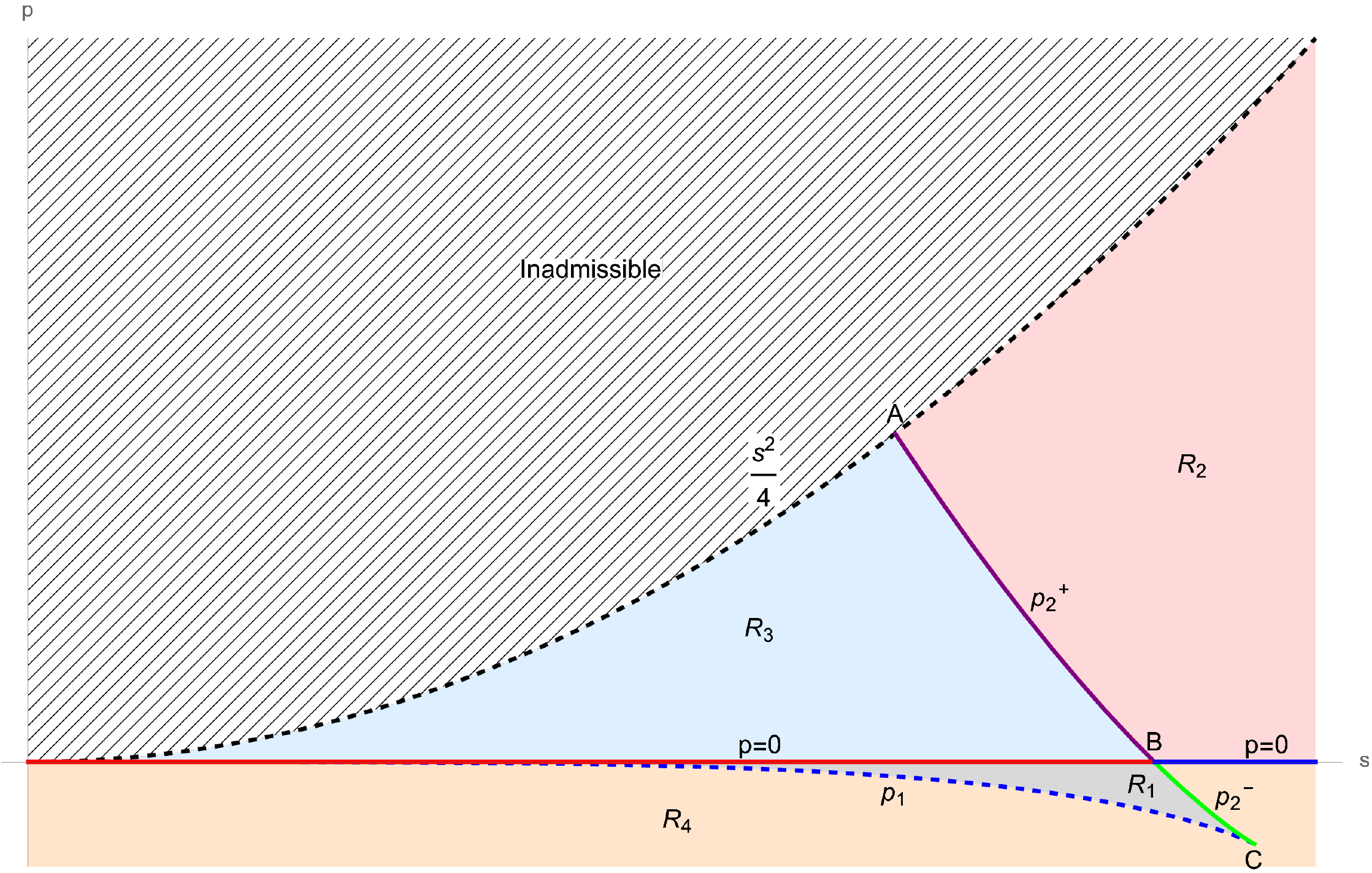}
%\captionsetup{width=0.75\linewidth}
\caption{A plot of the various domains of validity for solutions in $(s,p)$ space. The three coloured regions denoted $R_{1,2,3}$ allow for a global completion to a compact space, while region 4 allows for non-compact domain-wall solutions. The regions are separated by various lines, the non-trivial lines $p=p_1(s)$ and $p=p_2(s)$ are given explicitly in \eqref{eq:p1} and \eqref{eq:p2}. We have added a $\pm$ superscript on $p_2$ to denote whether it is above or below the axis, the form of $p_2$ is independent of this. There are three distinguished points arising from the intersection of the boundary lines: $A=\big(\tfrac{8}{27},\tfrac{16}{729}\big)$, $B=\big(\tfrac{2}{3\sqrt{3}},0\big)$ and $C=\big(\frac{8 \sqrt{2}}{27},-\tfrac{4}{729}\big)$.}
  \label{fig:psdomain}
\end{figure}

\paragraph{General global analysis} The global properties of the metric depend on the number of roots of the function, $F(y)$. To analyse this, let us define
\be
s=m^3 (q_1+q_2)\, ,\qquad p=m^6 q_1 q_2\,, \qquad \Leftrightarrow \qquad q_1=\frac{s-\sqrt{s^2-4 p}}{2 m^3}\, ,\quad q_2=\frac{s+\sqrt{s^2-4p}}{2m^3}\, .
\ee 
Clearly, for the $q_{i}$'s to be real we must impose that $s^2-4p\geq0$, with equality iff $q_1=q_2$.
Then, with these variables (and setting $g,m$ to their supersymmetric value, $2g=3m$) we have
\be
\hat{F}(y)\equiv m^4 F( y m^{-1})= p+y^3\big(y^3-y+s\big)\, ,
\ee
and it remains to study the roots of the above function while varying the two parameters $p,s$. Note that while the metric and gauge fields are invariant under the simultaneous transformation, $y\rightarrow -y, q_i\rightarrow-q_i$, the scalar fields are not and pick up a minus sign. We must restrict to the $p,s$ parameter space where $\hat{F}(y)>0$ and $h_i(y)>0$ in the domain of $y$ in order for the scalar fields and metric to be well-defined. From the form of the function one finds that there can be at most four real non-zero roots. If none of the roots are zero then there is necessarily a complex pair of roots. Furthermore, it is not hard to see that there is a root at zero if and only if one of the $q_i=0$, therefore, away from $p=0$, there is always a complex pair of roots. 
The form of the function, $\hat{F}(y)$, also implies that there are always two real roots, with a further two real roots appearing in a restricted domain. In figure \ref{fig:psdomain} we have plotted the various regions in $(p,s)$ parameter space giving rise to globally well-defined solutions. 

The regions are defined by four lines, which are further broken up by studying their intersections. The inadmissible region has $4p>s^2$ leading to complex $q_i$. Region 1 is has $p<0$ and, therefore, $q_i$'s are of opposite sign. In region 1 there are four real roots, three of which are positive. In regions 2 and 3 we have $p>0$, and so the $q_i$ are of the same sign. In region 2 there are only two real roots both of which are negative. In region 3 there are again four real roots, this time two positive and two negative. Finally in region 4 we find two real roots only, one positive and one negative. The two remaining lines are not as simple to write down but can be given explicitly by studying the vanishing of the discriminant of $\hat{F}(y)$. We find that the line bounding region 1 from below is given by
\be
p=p_1(s)\equiv\frac{1}{324}\Big(32+81 s^2 - 2^{1/3}\big( \omega (\alpha-\ii \beta)^{1/3}-\omega^2 (\alpha+\ii \beta)^{1/3}\big)\Big)\, ,\label{eq:p1}
\ee
where 
\be
\begin{split}
\omega=(-1)^{1/3}\, ,\quad \alpha= 27^2 s(320s+729 s^3 )-2048\, ,\quad \beta=27 s(128 -729 s^2)^{3/2}\, .
\end{split}
\ee
In the range, $0<s<\frac{8 \sqrt{2}}{27}$, this is real as it should be. The final line which forms the boundary between region 2 and 3 and caps off region 1 is
\be
p=p_2(s)\equiv \frac{1}{324}\Big(32+81 s^2 + 2^{1/3}\big( \omega^2 (\alpha-\ii \beta)^{1/3}-\omega (\alpha+\ii \beta)^{1/3}\big)\Big)\, ,\label{eq:p2}
\ee
again this is real for $0<s<\frac{8 \sqrt{2}}{27}$. Along these lines $\hat{F}(y)$ develops a double root and a triple root at the intersection point at $s=\frac{8 \sqrt{2}}{27}$, (point C in figure \ref{fig:psdomain}). 

Altogether we have five different end-points to consider\footnote{We will ignore the triple root here since we have not found a suitable way to incorporate this into a well-defined global completion, unless we take a scaling limit to the torus.}: a non-zero single root, a non-zero double root, a root at 0, 0 where $\hat{F}(0)>0$ and $\infty$. We have summarised the various options in table \ref{tab:endpoints} and the local geometry of $M_2$ around that point. Our goal is now to glue two choices of boundary condition together provided that such a choice of the gluing is possible. This can be read off from figure, $i.e.$, there is a non-trivial domain in figure \ref{tab:endpoints} where the two choice of end-points are possible. The different choices of gluing are summarised in table \ref{tab:differentM2}. 

\begin{table}
\begin{center}
\begin{tabular}{|c|c|}
\hline 
Special point & $M_2$ at special point\\
\hline
\hline
Single root & $\mathbb{R}^2/\mathbb{Z}_k$\\
\hline
Double root & $\mathbb{H}^2/\mathbb{Z}_k$\\
\hline 
Root at 0 & $ \mathbb{R}\times S^1/\mathbb{Z}_k$\\
\hline
0 but not a root &$\mathbb{R}\times S^1/\mathbb{Z}_k$\\
\hline
$\infty$ & Asymptotically AdS$_6$\\
\hline
\end{tabular}
\end{center}
\caption{The various choices of end-points for the global completion of the metric on $M_2$. The roots at 0 or an end-point at 0 without being a root both lead to singularities in the six-dimensional metric, these can be interpreted as smeared branes. For the final choice the metric on AdS$_4\times M_2$ becomes asymptotically locally AdS$_6$ space. }
\label{tab:endpoints}
\end{table}

\begin{table}
\begin{center}
\begin{tabular}{|c||c|c|}
\hline 
$M_2$ & Root structure &  Domain of validity \\
\hline\hline
Spindle $\mathbbl{\Sigma}$ \cite{Faedo:2021nub}& Two positive single roots &  Region 1  \\
\hline 
Disc $\mathbb{D}$ \cite{Suh:2021aik} & $\pm$ Single root and 0 & Region 2 and 3\\
\hline
Disc $\mathbb{D}$ (E.S.) (here) & Single root and root at 0 & $p=0$, $s\in (0,\tfrac{2}{\sqrt{27}})$
\\
\hline
Black Bottle $\mathbb{BB}$  (here)& positive single root and double root & $p=p_2^{-}$, $s\in\big(\tfrac{2}{3\sqrt{3}},\frac{8\sqrt{2}}{27}\big)$\\
\hline
Black Goblet $\mathbb{BG}$  (here)& 0 and double root & $p=p_2^+$, $s\in \big(\tfrac{8}{27},\tfrac{2}{3\sqrt{3}}\big)$\\
\hline
$S^2$ \cite{Karndumri:2015eta}& Scaling limit at double root& $p=p_1$, $s\in (0,\frac{8\sqrt{2}}{27})$ \\
\hline
$T^2$ (here)& Scaling limit at triple root& Point C \\
\hline
{$\mathbb{H}^2$} \cite{Karndumri:2015eta}& Scaling limit at double root & $p=p_2 $, $s\in\big(\tfrac{8}{27},\tfrac{8\sqrt{2}}{27}\big)$\\
\hline
$\mathbb{H}^2$ (E.S.)\cite{Karndumri:2015eta} & Scaling limit at double root& Point B\\
\hline\hline
Domain wall: conical & positive single root and $\infty$ & Region 1,3,4\\
\hline
Domain wall: disc  & 0 and $\infty$ & Region 2\\ \hline
Domain wall: disc  (E.S.) & root at 0 and $\infty$ & $p=0$, $s\in (\frac{2}{3\sqrt{3}},\infty)$\\ \hline
Domain wall: bottle & positive double root and $\infty$ & $p=p_2$, $s\in \big(\tfrac{8}{27},\tfrac{8\sqrt{2}}{27}\big)$\\\hline
AdS$_6$ \cite{Romans:1985tw, Brandhuber:1999np} & –& $p=s=0$\\
\hline
\end{tabular}
\end{center}
\caption{The different choices of global completion and the values of $p,s$ giving rise to such solutions. We have indicated where these solutions were first considered, in some cases it is in this paper. Certain solutions lead to an enhanced symmetry group in the uplifted solution and have been distinguished by writing E.S. (enhanced symmetry). The solutions of the first block have finite free energies when viewed as an AdS$_4$ solution. The second block of solutions have infinite free energies due to the AdS$_6$ asymptotics. }
\label{tab:differentM2}
\end{table}

We spare the reader the details of studying all the solutions appearing in table \ref{tab:differentM2} here and refer them to appendix \ref{app:globalsols} for this analysis. Here we will present only the analysis of spindle since this will be the main focus of our consistent truncation, though we emphasise that it works for any of the solutions given in table \ref{tab:differentM2}.

\paragraph{Spindle} Here we consider the spindle completion of the local metric above, as originally discussed in \cite{Faedo:2021nub}. In order for the metric to give rise to a spindle the parameters must be chosen to lie in region $1$ of figure \ref{fig:psdomain} so that there are positive non-zero single roots at both end-points. The full domain is given by
\be
s\in \big(0,\tfrac{8 \sqrt{2}}{27}\big)\qquad \text{ and}\qquad  p_1(s)< p< \text{min}\big(0,p_2^{-}(s)\big) \, ,
\ee
where the bounds on $p$ are $s$-dependent. Within this region, and away from the boundaries, we have three real positive roots and one negative root. Let $y_-$ be the smallest positive root and $y_+$ the next positive root. The local metric can be completed to a well-defined global metric on a spindle, \cite{Faedo:2021nub}, by taking the domain of $y$ to be $y \in [y_-,y_+]$. The two end-points correspond to the two poles of the spindle which have conical singularities, $\mathbb{R}^2/\mathbb{Z}_{n_\pm}$. The orbifold weights, $n_+$, $n_-$, are taken to be relatively prime. The period of the circle direction is
\begin{equation}
\frac{\Delta z}{2\pi}=\frac{2 y_{+}^3}{|F'(y_{+})|m n_{+}}=\frac{2 y_{-}^3}{|F'(y_{-})|m n_{-}}\, .
\end{equation}

Using the metric we can compute the Euler characteristic of the spindle, finding
\begin{equation}
\chi\left(\mathbbl{\Sigma}_2\right)\,=\,\frac{1}{4\pi}\int_{\mathbbl{\Sigma}_2}R_{\mathbbl{\Sigma}_2}\text{vol}_{\mathbbl{\Sigma}_2}\,=\,\frac{n_++n_-}{n_+n_-}\,.
\end{equation}
The magnetic flux threading through the spindle is quantised according to
\begin{equation}
\frac{g}{2\pi}\int_{\mathbbl{\Sigma}_2}F_i\, =\frac{g\Delta z}{2\pi}\bigg( \frac{y_{-}^3}{h_i(y_-)}-\frac{y_{+}^3}{h_i(y_+)}\bigg)\,\equiv\,\frac{p_i}{n_+n_-}\,,
\end{equation}
and the integer flux numbers, $p_1$, $p_2$, satisfy the twist condition,
\begin{equation}
p_1+p_2\,=\,n_-+n_+\,,
\end{equation}
upon application of $3m=2g$.
Note that the graviphoton field strength, $F^R$, is given by
\be
F^R=g (F_1+F_2)\, ,
\ee
and, therefore, the solution realises the ``twist" mechanism for preserving supersymmetry.

One may express the quantities, $y_\pm$, $q_1$, $q_2$ and $\Delta{z}$, in terms of the orbifold weights, $n_+$, $n_-$, and the flux numbers, $p_1$, $p_2$. Following \cite{Faedo:2021nub} we reparametrise $y_\pm$ in terms of $w$ and $\tau$ as
\begin{equation}
y_-\,=\,w\left(1-\tau\right)\,, \qquad y_+\,=\,w\left(1+\tau\right)\,.
\end{equation}
We also introduce $\mu$ and ${\tt z}$  by
\begin{equation}
\mu\,=\,\frac{n_+-n_-}{n_++n_-}\,,
\end{equation}
and
\begin{equation}
p_1\,=\,\frac{n_++n_-}{2}\left(1+{\tt z}\right)\,, \qquad p_2\,=\,\frac{n_++n_-}{2}\left(1-{\tt z}\right)\,.
\end{equation}
By solving the following quartic equation,
\begin{equation} \label{polyp}
P(\tau)\,=\,\tau^4+\left(8{\tt z}^2-3-9\mu^2\right)\tau^2+12\mu{\tau}-9\mu^2\,=\,0\,,
\end{equation}
$\tau$ can be expressed in terms of the orbifold weights, $n_+$, $n_-$, and the flux numbers, $p_1$, $p_2$.
It follows that we can express $y_\pm$, $q_1$, $q_2$, $w$ and $\Delta{z}$ in terms of $\tau$, $\mu$ and ${\tt z}$, 
\begin{align} \label{yns}
y_{\pm}\,=&\,\frac{1\pm{\tau}}{g\left(\tau^2+3\right)}\sqrt{\frac{9\mu\left(\tau^2+1\right)-3\tau\left(\tau^2+5\right)}{2\left(\mu-\tau\right)}}\,, \\ \label{q12}
q_{1,2}\,=&\,w\frac{3\left(1-
\tau^2\right)}{g^2\left(\tau^2+3\right)^2\left(\mu-\tau\right)}\left[3\mu\left(1+\tau^2\right)-2\tau\mp{\tau}\left(\tau^2+3\right){\tt z}\right]\,, \\ \label{wdef}
w\,=&\,\frac{1}{g\left(\tau^2+3\right)}\sqrt{\frac{9\mu\left(\tau^2+1\right)-3\tau\left(\tau^2+5\right)}{2\left(\mu-\tau\right)}}\,, \\ \label{delz}
\Delta{z}\,=&\,\chi(\mathbbl{\Sigma}_2)\frac{3\pi\left(\tau^2+3\right)\left(\mu-\tau\right)}{8g\tau^2}\,.
\end{align}

The holographic dual of the solution should be a 3d $\mathcal{N}=2$ SCFT obtained by compactifying 5d $\mathcal{N}=1$ SCFT with gauge group USp(2$N$), \cite{Seiberg:1996bd}, on the spindle, $\mathbbl{\Sigma}_2$. The explicit details of this 3d SCFT are currently lacking, however. Using the uplift formulae provided in \cite{Faedo:2021nub}, and extended in appendix \ref{app:Uplift}, one may uplift this solution to massive type IIA supergravity.\footnote{One could also consider uplifting this to type IIB supergravity, see for example \cite{Malek:2019ucd} which considers the embedding of F$(4)$ gauged supergravity coupled to vector multiplets into type IIB supergravity and the embedding of pure F$(4)$ gauged supergravity in type IIB supergravity in \cite{Jeong:2013jfc, Hong:2018amk, Malek:2018zcz}. The analysis above shows that for the spindle it is necessary to consider the theory coupled to a vector multiplet, not just the pure theory, so there is currently no known consistent truncation which we may work with.} The holographic free energy of the solution is
\begin{align} \label{s3sig2free}
\mathcal{F}_{S^3\times{\mathbbl{\Sigma}_2}}\,=&\,\frac{16\pi^3l^8}{\left(2\pi{l}_s\right)^8}\frac{3\pi^2\lambda^4}{10g^4}\left(y_+^3-y_-^3\right)\Delta{z} \\
=&\,\chi(\mathbbl{\Sigma}_2)\frac{\sqrt{3}\pi{N}^{5/2}}{5\sqrt{8-N_f}}\frac{\left[3\mu\left(\tau^2+1\right)-\tau\left(\tau^2+5\right)\right]^{3/2}}{\tau\left(\tau^2+3\right)\left(\mu-\tau\right)^{1/2}}\,,
\end{align}
where $N$ and $N_f$ are the number of D4- and D8-branes, respectively.

%%%%%%%%%%%%%%%%%%%%%%%%%%%%%%%%%%%%%
\subsection{Consistent truncation}
%%%%%%%%%%%%%%%%%%%%%%%%%%%%%%%%%%%%%

Using the AdS$_4\times M_2$ solution given in \eqref{ads4sigma2} as our seed geometry, we construct a consistent truncation of matter coupled $F(4)$ gauged supergravity on a spindle, $\mathbbl{\Sigma}_2$, to four-dimensional minimal gauged supergravity. After a little trial and error, we find that the ansatz for the metric is
\begin{equation} \label{upansatz1}
ds_6^2\,=\,\left(y^2h_1(y)h_2(y)\right)^{1/4}\left[ds_4^2+\frac{y^2}{F(y)}dy^2+\frac{F(y)}{h_1(y)h_2(y)}\left(dz-\frac{1}{2m}\mathcal{A}\right)^2\right]\,,
\end{equation}
where $ds_4^2$ and $\mathcal{A}$ are the metric and the gauge field of four-dimensional minimal gauged supergravity. Accordingly, the gauge fields and the two-form field are given by
\begin{align} \label{upansatz2}
A_i\,=&\,-\frac{y^3}{h_i(y)}\left(dz-\frac{1}{2m}\mathcal{A}\right)\,, \notag \\
B\,=&-\,\frac{1}{m}y*_4\mathcal{F}\,,
\end{align}
where $\mathcal{F}=d\mathcal{A}$. 
The scalar fields are the same as the ones of the AdS$_4\times\mathbbl{\Sigma}_2$ solutions,
\begin{equation} \label{upansatz3}
X_i\,=\,\left(y^2h_1(y)h_2(y)\right)^{3/8}h_i(y)^{-1}\,.
\end{equation}
Note that we have set the constant gauge parameter introduced in the gauge fields in \eqref{ads4sigma2} to zero. That is, by introducing the graviphoton via minimal coupling, $d z \rightarrow d z-\tfrac{1}{2m}\mathcal{A}$, one must choose the gauge, $\alpha_i=0$. Given that, in the uplifted theory, the U$(1)$ isometry of the spindle mixes with the R-symmetry and this is not too unexpected. In the uplifted metric different gauges for the two gauge fields correspond to different choices of the R-symmetry direction. It is precisely the gauge, $\alpha_i=0$, that in the uplifted solution the R-symmetry direction is $\partial_{z}$, and any other gauge leads to an R-symmetry direction which mixes the $z$ coordinate with those of the isometry directions of the four-dimensional hemi-sphere in the uplift. Since the graviphoton is the gauge field of the R-symmetry, it should appear by gauging the exact R-symmetry direction. Thus, if this is by a minimal coupling term, the gauge choice is important and, more importantly, fixed to the one we consider. A similar argument also explains why \cite{Cheung:2022ilc} found the need for a particular gauge when considering the truncation of seven-dimensional gauged supergravity on a spindle to five-dimensional minimal gauged supergravity.

If we substitute the ansatz to the equations of motion of $F(4)$ gauged supergravity, they reduce to the equations of motion of four-dimensional minimal gauged supergravity,
\begin{align}
R_{\alpha\beta}\,&=\,-3g_{\alpha\beta}+\frac{1}{2}\left(\mathcal{F}_{\alpha\gamma}\mathcal{F}_\beta\,^\gamma-\frac{1}{4}g_{\alpha\beta}\mathcal{F}_{\gamma\delta}\mathcal{F}^{\gamma\delta}\right)\,, \notag \\
\dd\star_4 \mathcal{F}&=\,0\,.
\end{align}
The equations of motion can be derived from the action,
\begin{equation}
S\,=\,\frac{1}{16\pi{G}_N^{(4)}}\int{d}x^4\sqrt{-g}\left(R+6-\frac{1}{4}\mathcal{F}_{\alpha\beta}\mathcal{F}^{\alpha\beta}\right)\,,
\end{equation}
where the cosmological constant was normalised so that $L_{AdS_4}=1$.

The truncation we have presented above is a local construction and indeed no data about the global completion of the solution was used in deriving the truncation. Thus, it can equally be applied to construct truncations on the other backgrounds that we study in appendix \ref{app:globalsols}.\footnote{One subtlety arises for the constant curvature Riemann surfaces. These are obtained by taking a certain scaling limit of the solutions and in this limit the gauging of the isometry direction in the metric on $M_2$ vanishes. The presence of the gauge field of four-dimensional minimal gauged supergravity is still present in the flux terms, however, so this does still give a sensible consistent truncation in these cases. This truncation is more in the spirit of \cite{Faedo:2019cvr,Hosseini:2020wag}.}

%%%%%%%%%%%%%%%%%%%%%%%%%%%%%%%%%%%%%
\section{AdS$_2\times\mathbbl{\Sigma}_1\ltimes\mathbbl{\Sigma}_2$ solutions}\label{sec:spinspin}
%%%%%%%%%%%%%%%%%%%%%%%%%%%%%%%%%%%%%

Employing the consistent truncation of the previous section, in particular, \eqref{upansatz1}, \eqref{upansatz2}, and \eqref{upansatz3}, we will uplift the known spindle solution, AdS$_2\times\mathbbl{\Sigma}_1$, in four-dimensional minimal gauged supergravity to $F(4)$ gauged supergravity. We obtain a class of AdS$_2\times\mathbbl{\Sigma}_1\ltimes\mathbbl{\Sigma}_2$ solutions where the compact space is a four-dimensional orbifold. Unlike the AdS$_4\times\mathbbl{\Sigma}_2$ solutions which are in the twist class, the minimal AdS$_2\times\mathbbl{\Sigma}_1$ solutions preserve supersymmetry via the anti-twist. It would be interesting to extend our consistent truncation beyond the minimal case studied here. Solutions in four-dimensional U$(1)^4$ gauged supergravity have been constructed in \cite{Ferrero:2021ovq,Couzens:2021rlk}, including disc solutions, and uplifting solutions in these large four-dimensional theories would allow for us to uplift solutions which preserve supersymmetry by the twist instead, \cite{Ferrero:2021etw,Couzens:2021cpk}. 

%%%%%%%%%%%%%%%%%%%%%%%%%%%%%%%%%%%%%
\subsection{Uplifting $D=4$ to $D=6$}
%%%%%%%%%%%%%%%%%%%%%%%%%%%%%%%%%%%%%

In four-dimensional minimal gauged supergravity, supersymmetric AdS$_2\times\mathbbl{\Sigma}_1$ solutions were constructed in \cite{Ferrero:2020twa}.\footnote{In \cite{Ferrero:2020twa} these solutions were allowed to rotate. In the following we will restrict to the static case for simplicity of exposition, though much of the analysis for the rotating case is the same as that presented here. This would constitute the first example of a rotating four-dimensional orbifold albeit in only one plane rather than the two possible planes.} The metric and the gauge field are
\begin{equation}
ds_4^2\,=\,\frac{x^2}{4}ds_{\text{AdS}_2}^2+ds_{\mathbbl{\Sigma}_1}^2\,, \qquad \mathcal{A}\,=\,\left(1-\frac{\mathtt{a}}{x}\right)d\psi\,,
\end{equation}
where the metric on the spindle, $\mathbbl{\Sigma}_1$, is 
\begin{equation}
ds_{\mathbbl{\Sigma}_1}^2\,=\,\frac{x^2}{q(x)}dx^2+\frac{q(x)}{4x^2}d\psi^2\,,
\end{equation}
and the function, $q(x)$, is 
\begin{equation}
q(x)\,=\,x^4-4x^2+4\mathtt{a}x-\mathtt{a}^2= x^4 - (2x-\mathtt{a})^2\,.\label{eq:q(x)def}
\end{equation}

This local metric can be extended to a metric on a spindle, with relatively prime orbifold weights, $m_+$ and $m_-$, and a suitably quantised magnetic flux provided the period, $\Delta{\psi}$, and the parameter, $\mathtt{a}$, are fixed to be
\begin{align} \label{xxxx1}
\mathtt{a}\,=&\,\frac{m_+^2-m_-^2}{m_+^2+m_-^2}\,, \notag \\
\Delta\psi\,=&\,\pi\sqrt{2}\frac{\sqrt{m_+^2+m_-^2}}{m_+m_-}\,.
\end{align}
The line interval with coordinate $x$ is fixed to lie in $x\in[x_-,x_+]$ where $x_-$, $x_+$ are the two middle roots of $q(x)$ and take the form,
\begin{align} \label{xxxx2}
x_-\,=&\,-1+\sqrt{1+\mathtt{a}}\,, \notag \\
x_+\,=&\,1-\sqrt{1-\mathtt{a}}\,.
\end{align}
The Euler characteristic of the spindle is given by
\begin{equation}
\chi\left(\mathbbl{\Sigma}_1\right)\,=\,\frac{1}{4\pi}\int_{\mathbbl{\Sigma}_1}R_{\mathbbl{\Sigma}_1}\text{vol}_{\mathbbl{\Sigma}_1}\,=\,\frac{m_++m_-}{m_-m_+}\,,
\end{equation}
and the magnetic flux through the spindle is
\begin{equation} \label{intf2four}
\frac{1}{2\pi}\int_{\mathbbl{\Sigma}_1}\mathcal{F}\,=\,\frac{m_+-m_-}{m_-m_+}\,.
\end{equation}
From \eqref{intf2four} we see that the solutions are in the anti-twist class after noting that the gauge field for the R-symmetry is $\mathcal{A}^{R}\equiv\mathcal{A}$ . Finally the Bekenstein--Hawking entropy is given by
\begin{equation} \label{4dbharea}
S_{\text{BH}}\,=\,\frac{A_h}{4G_N^{(4)}}\,=\,\frac{\pi}{4G_N^{(4)}}\frac{\sqrt{2}\sqrt{m_-^2+m_+^2}-\left(m_-+m_+\right)}{m_-m_+}\,,
\end{equation}
where $A_h\,=\,\frac{1}{2}\left(x_+-x_-\right)\Delta{\psi}$ is the area of the horizon.

By employing the consistent truncation ansatz worked out earlier in \eqref{upansatz1}, \eqref{upansatz2} and \eqref{upansatz3}, we can uplift the AdS$_2\times\mathbbl{\Sigma}_1$ solution to $F(4)$ gauged supergravity and obtain a class of AdS$_2\times\mathbbl{\Sigma}_1\ltimes\mathbbl{\Sigma}_2$ solutions given by,
\begin{align}
ds_6^2\,=&\,\left(y^2h_1(y)h_2(y)\right)^{1/4}\left[\frac{x^2}{4}ds_{\text{AdS}_2}^2+\frac{x^2}{q(x)}dx^2+\frac{q(x)}{4x^2}d\psi^2\right. \notag \\
& \,\,\,\,\,\,\,\,\,\,\,\,\,\,\,\,\,\,\,\,\,\,\,\,\,\,\,\,\,\,\,\,\,\,\,\,\,\,\,\,\,\,\,\,\,\,\,\,\,\,\,\,\,\,\,\,\,\,\, \left.+\frac{y^2}{F(y)}dy^2+\frac{F(y)}{h_1(y)h_2(y)}\left(dz-\frac{1}{2m}\left(1-\frac{\mathtt{a}}{x}\right)d\psi\right)^2\right]\,, \notag \\
A_i\,=&\,-\frac{y^3}{h_i(y)}\left(dz-\frac{1}{2m}\left(1-\frac{\mathtt{a}}{x}\right)d\psi\right)\,, \notag \\
B\,=&\,- \frac{\mathtt{a}}{2m}y\,\text{vol}_{\text{AdS}_2}\,, \notag \\
X_i\,=&\,\left(y^2h_1(y)h_2(y)\right)^{3/8}h_i(y)^{-1}\,.
\end{align}

From the form of the metric we see that the spindle, $\mathbbl{\Sigma}_2$, is non-trivially fibered over the spindle, $\mathbbl{\Sigma}_1$. In order for this fibration to be well-defined, the one-form, $\eta=\frac{2\pi}{\Delta{z}}\left(dz-\frac{1}{2m}\left(1-\frac{\mathtt{a}}{x}\right)d\psi\right)$, should be globally defined. This is equivalent to requiring that the curvature of the bundle over $\mathbbl{\Sigma}_1$ is in $\mathbb{Z}/m_-m_+$, that is we require
\begin{equation}
\frac{1}{2\pi}\int_{\mathbbl{\Sigma}_1}d\eta\,=\,\frac{t}{m_-m_+}\,, \qquad t\in\mathbb{Z}\,.
\end{equation}
Since $d\eta=-\frac{2\pi}{\Delta{z}}\frac{1}{2m}\mathcal{F}$, using \eqref{delz} and \eqref{intf2four}, we find a condition relating the spindle numbers, $m_\pm$, $n_\pm$, and the flux numbers, $p_1$, $p_2$, 
\begin{equation}\label{eq:tdef}
t\,=\,\left(m_+-m_-\right)\frac{n_+n_-}{n_++n_-}\frac{4\tau^2}{\left(
\tau^2+3\right)\left(\mu-\tau\right)}\,\in\,\mathbb{Z}\,.
\end{equation}
This condition ensures that away from the poles on the $\mathbbl{\Sigma}_2$ fibre, the space $\mathbbl{\Sigma}_1\ltimes\mathbbl{\Sigma}_2$ is well-defined. In fact the circle parametrised by $z$ fibered over the spindle, $\mathbbl{\Sigma}_1$, gives the Lens space $L(t,1)$. While the conical singularities of $\mathbbl{\Sigma}_1$ are removed, those of $\mathbbl{\Sigma}_2$ remain. This is an orbifold version of a Hirzebruch surface where the $\mathbb{CP}^1$'s have been replaced with spindles, \cite{Cheung:2022ilc}. As a final comment it is possible to find solutions of the parameters such that this quantisation condition is satisfied, thus the set of well-defined solutions is non-empty. 

%%%%%%%%%%%%%%%%%%%%%%%%%%%%%%%%%%%%%
\subsection{Uplifting to massive type IIA supergravity}
%%%%%%%%%%%%%%%%%%%%%%%%%%%%%%%%%%%%%

We now want to consider uplifting the solution to massive type IIA supergravity on the Brandhuber--Oz solution, $i.e.$, with internal manifold a four-dimensional hemisphere. The uplift of six-dimensional U$(1)^2$-gauged supergravity, where the two-form potential, $B$, is set to vanish, was given in \cite{Faedo:2021nub} with the gauge coupling and Romans mass fixed to satisfy $3m=2g$. In appendix \ref{app:Uplift} we have constructed the general uplift allowing for a non-trivial two-form field, $B$, and also allowing for the Romans mass and gauge coupling to be independent. Despite the generality of our uplift, in the following, we will impose the supersymmetry constraint, $3m=2g$, since we are interested in supersymmetric solutions of massive type IIA supergravity.

In appendix \ref{app:Uplift} we have introduced two free parameters, $\lambda$ and $l$, to the uplift formula. The parameter, $l$, is an overall length scale and can be thought of as the length scale of the background AdS$_6$ solution. The other parameter, $\lambda$, is less intuitive but is required for flux quantisation. Both are symmetries of the action, but only $l$ is a symmetry for flux quantisation since it may be absorbed by redefining the string length. The metric in string frame reads
\begin{align}
ds_{\text{s.f.}}^2\,&=\,\lambda^2l^2\mu_0^{-1/3}\left\{y^{-1}\Delta_h^{1/2}\widehat{ds_6^2}+g^{-2}\Delta^{-1/2}_h\left[y^3d\mu_0^2+\sum_{i=1}^{2}h_i(y)\left(d\mu_i^2+\mu_i^2D\phi_i^2\right)\right]\right\}\,,
\end{align}
where we have defined
\begin{align}
\widehat{ds_6^2}\,=\,\left[\frac{x^2}{4}\right.ds_{\text{AdS}_2}^2&+\frac{x^2}{q(x)}dx^2+\frac{q(x)}{4x^2}d\psi^2 \left.+\frac{y^2}{F(y)}dy^2+\frac{F(y)}{h_1(y)h_2(y)}\left(dz-\frac{1}{2m}\left(1-\frac{\mathtt{a}}{x}\right)d\psi\right)^2\right]\,,
\end{align}
and the one-forms, $D\phi_i=d\phi_i-gA_i$. The angular coordinates, $\phi_i$, have the canonical period $2\pi$. 
The functions appearing in the metric are
\begin{align}
\Delta_h\,=&\,h_1(y)h_2(y)\mu_0^2+y^3h_2(y)\mu_1^2+y^3h_1(y)\mu_2^2\,, \notag \\
U_h\,=&\,2\left[\left(y^3-h_1(y)\right)\left(y^3-h_2(y)\right)\mu_0^2-y^6\right]-\frac{4}{3}\Delta_h\,.
\end{align}

The solution is supported by the full complement of fields: there are non-trivial NSNS two-form potential, $B_{(2)}$, RR one- and three-form potentials, $C_{(1)}$ and $C_{(3)}$, dilaton field, $\Phi$, and the Romans mass, $F_{(0)}$. The field strengths of the potentials are given by
\begin{equation}
H_{(3)}\,=\,dB_{(2)}\,, \quad F_{(2)}\,=\,dC_{(1)}+F_{(0)}B_{(2)}\,, \quad F_{(4)}\,=\,dC_{(3)}+B_{(2)}\wedge F_{(2)}-\frac{1}{2}F_{(0)} B_{(2)}\wedge B_{(2)}\,.
\end{equation}
In particular, the two-form flux and the NSNS two-form potential are given, respectively, by
\begin{align}
\lambda l^{-1}{F}_{(2)}\,=&\,-\frac{m}{2}\mu_0^{2/3}B\,, \notag \\
\lambda^{-2}l^{-2}B_{(2)}\,=&\,-\frac{1}{2}\mu_0^{2/3}B\,,
\end{align}
where $B$ is the two-form field of $F(4)$ gauged supergravity. Also the dilaton field and the Romans mass are given by
\begin{equation}
e^\Phi\,=\,\lambda^2\mu_0^{-5/6}y^{-3/2}\Delta^{1/4}_h\,,
\end{equation}
\begin{equation}
F_{(0)}\,=\,\frac{m}{\lambda^3 l}\,.
\end{equation}
Due to the unwieldy expression of the four-form flux, $F_{(4)}$ we present only the terms that are pertinent for flux quantisation: these are the terms on either the hemi-sphere or on the four-dimensional orbifold $\mathbbl{\Sigma}_1\ltimes\mathbbl{\Sigma}_2$,
\begin{align} \label{eq:F4exp}
\lambda^{-1}l^{-3}F_{(4)}\,=&\,\frac{\mu_0^{1/3}h_1h_2}{g^3\Delta_h}\frac{U_h}{\Delta_h}\frac{\mu_1\mu_2}{\mu_0}d\mu_1\wedge{d}\mu_2\wedge{D}\phi_1\wedge{D}\phi_2-\frac{g}{3}\frac{\mu_0^{4/3}}{X_1X_2}\star_6B+\ldots\,.
\end{align}
The remaining terms can be read off from appendix \ref{app:Uplift} after a little work by the reader.

Let us define the two-cycles, $S_a\equiv\{y=y_a\}$, $a=\pm$, to be the two sections defined at the two poles of the fibre, $\mathbbl{\Sigma}_2$, \cite{Cheung:2022ilc}, respectively. The fluxes through the two-cycles are
\begin{align}
\frac{g}{2\pi}\int_{S_+}F_i\,=&\,\frac{g}{2m}\frac{y_+^3}{y_+^3+q_i}\frac{m_--m_+}{m_-m_+}\,, \notag \\
\frac{g}{2\pi}\int_{S_-}F_i\,=&\,\frac{g}{2m}\frac{y_-^3}{y_-^3+q_i}\frac{m_--m_+}{m_-m_+}\,.
\end{align}
Recall that the R-symmetry field strength is $F^R=g(F_1+F_2)$. Then, on the two-cycles, $S_a$, we find
\begin{align}
\frac{1}{2\pi} \int_{S_+} F^R&= \frac{m_+-m_-}{m_- m_+}+\frac{t}{n_+ m_- m_+}\, ,\nonumber \\
\frac{1}{2\pi} \int_{S_-} F^R&=\frac{m_+-m_-}{m_-m_+}-\frac{t}{n_- m_-m_+}\, .\label{eq:FRoverS}
\end{align}
Note that we have the homology relation, $S_+-S_-\,=\,\frac{t}{m_-m_+}\mathbbl{\Sigma}_2$, (see \cite{Cheung:2022ilc} for the analogous relation in the M5-brane case), and indeed this holds for the R-symmetry field strength,
\begin{equation}
\frac{1}{2\pi}\int_{S_+}F^R-\frac{1}{2\pi}\int_{S_-}F^R=\frac{t}{m_-m_+} \frac{1}{2\pi}\int_{\mathbb{\Sigma}_2}F^R\, .
\end{equation}
The R-symmetry flux through the spindle, $\mathbbl{\Sigma}_2$, is
\begin{equation}
\frac{1}{2\pi}\int_{\mathbbl{\Sigma}_2}F^R\,=\,\frac{1}{n_-}+\frac{1}{n_+}\,=\frac{1}{2\pi}\,\int_{\mathbbl{\Sigma}_2}c_1\left(\mathbbl{\Sigma}_2\right)\,,
\end{equation}
where $c_1(\mathbbl{\Sigma}_2)$ is the first Chern class of the spindle. 

For the solution to be well-defined in string theory we must quantise the fluxes appropriately. There are two quantization conditions on the fluxes which are inherited from the AdS$_6$ vacuum solution, giving the integers, $N$ and $N_f$, of the parent theory,
\begin{equation} \label{fluxq}
\left(2\pi{l}_s\right)F_{(0)}\,\equiv\,n_0\,\in\,\mathbb{Z}\,, \qquad \frac{1}{\left(2\pi{l}_s\right)^3}\int_{\hat{S}^4}F_{(4)}\,\equiv\,N\,\in\,\mathbb{Z}\,,
\end{equation}
where the second one is the four-form flux through the four-hemisphere. We denote by $\ell_s$ the string length and the number of D8-branes, $N_f$, is fixed in terms of the flux quantum, $n_0$, according to $n_0=8-N_f$. For the solution, these imply the quantisation conditions
\begin{equation} \label{fluxquanc}
g^8\,=\,\frac{l^{8}}{\left(2\pi{\ell}_s\right)^8}\frac{18\pi^6}{N^3n_0}\,, \qquad \lambda^8\,=\,\frac{8\pi^2}{9Nn_0^3}\,.
\end{equation}
There is one further four-cycle upon which we may quantise the fluxes, namely the orbifold $M_4$ which appears in the title of this paper. This is constructed by going to the pole of the hemi-sphere at $\mu_0=1$, giving a copy of the four-dimensional orbifold. The four-form flux through the orbifold four-cycle, $\mathbbl{\Sigma}_1\ltimes\mathbbl{\Sigma}_2$, is
\begin{align}
\frac{1}{\left(2\pi{\ell}_s\right)^3}\int_{\mathbbl{\Sigma}_1\ltimes\mathbbl{\Sigma}_2}F_{(4)}\,&=\,\frac{1}{6\pi}\frac{m_+-m_-}{m_-m_+}\left[g^2\left(y_-^2-y_+^2\right)\right]\big[g\Delta{z}\big]N\nonumber\\
&=\frac{m_+-m_-}{m_-m_+}\frac{n_+ +n_-}{n_- n_+}\frac{3 \mu(\tau^2+1)- \tau(\tau^2+5)}{ 8 \tau (\tau^2+3)}N\,,
\end{align}
where we have used \eqref{yns}, \eqref{delz}, and \eqref{fluxquanc}. Note that the quantities, $\left[g^2\left(y_+^2-y_-^2\right)\right]$ and $\left[g\Delta{z}\right]$, are independent of the gauge coupling, $g$. It is an open question what quantization condition we should impose on this quantity and indeed this was also a problem in the M5-brane case, \cite{Cheung:2022ilc}. Nevertheless, we may arrange for this to be integer by tuning $N$. However, this seems too restrictive and a weaker condition should probably be imposed.

This AdS$_2\times\mathbbl{\Sigma}_1\ltimes\mathbbl{\Sigma}_2\ltimes \hat{S}^4$ geometry should correspond to the near-horizon of a black hole, as such we may evaluate the leading order correction to the Bekenstein--Hawking entropy. For a string frame metric of the form,
\begin{equation}
ds_{\text{s.f.}}^2\,=\,e^{2A}ds_{\text{AdS}_2}^2+ds_{M_8}^2\,,
\end{equation}
we have
\be
S_{\text{BH}}=\frac{1}{4 G_{N}^{(2)}}=\frac{1}{4G_{N}^{(10)}}\int_{M_8} \me^{-2\Phi}\text{vol}(M_8)\, ,
\ee
where the dilaton factor is because we are in string frame. For our solutions arising from the uplift of a four-dimensional black hole, the metric takes the factorised form,
\begin{equation}
ds^2= e^{2B} \left( e^{2C} \left(\dfrac{x^2}{4}ds_{\textrm{AdS}_2}^2 +ds^2_{N_2^{(1)}}+ds^2_{N_2^{(2)}}\right)+ds^2_{M_4}\right),
\end{equation}
where the warp factors are $e^{2B}=\lambda^2 l^2 \mu_0^{-1/3}(X_1 X_2)^{-1/4} \Delta^{1/2}$ and $e^{2C}=(y^2 h_1 h_2)^{1/4}$ and it is understood that $N_2^{(2)}$ is fibered over $N_2^{(1)}$.
Then the Bekenstein--Hawking entropy is
\begin{equation}
S_{\textrm{BH}}=\dfrac{1}{4G_N^{(2)}}= \dfrac{1}{4 G_N^{(10)}} \int e^{-2\Phi}e^{8B}e^{4C} \textrm{vol}(N_2^{(1)})\textrm{vol}(N_2^{(2)})\textrm{vol}(M_4)\, .
\end{equation}
To simplify this, recall that the holographic free energy of the AdS$_4$ solution, \eqref{s3sig2free}, is
\begin{equation}
\mathcal{F}_{S^3 \times N_2^{(2)}}=\dfrac{\pi}{2 G_N^{(4)}}=\dfrac{\pi}{2 G_N^{(10)}} \int e^{-2\Phi}e^{8B}e^{4C} \textrm{vol}(N_2^{(2)})\textrm{vol}(M_4).
\end{equation}
It follows that we may rewrite the Bekenstein--Hawking entropy in terms of the holographic free energy of the parent theory, \eqref{s3sig2free}, as
\begin{equation}
S_{\textrm{BH}}=\dfrac{A_h}{2\pi} \mathcal{F}_{S^3 \times N_2^{(2)}}\, ,\label{eq:SBHfactor}
\end{equation}
where $A_h$ is the area of the horizon of the four-dimensional black hole. By plugging the solution at hand in we find that the Bekenstein--Hawking entropy is
\be
S_{\text{BH}}=\frac{\sqrt{2}\sqrt{m_+^2+m_-^2}-\left(m_++m_-\right)}{2m_+m_-}\mathcal{F}_{S^3\times{\mathbbl{\Sigma}_2}}\, .
\ee
This factorised form of the entropy is a consequence, as is the dependence of the parameter, $t$ in \eqref{eq:tdef} on the spindle weights, of using our truncation to construct the four-dimensional orbifold horizons: one naturally has the splitting of the integral in \eqref{eq:SBHfactor} which is a generic feature of the uplift here. More general four-dimensional orbifolds, outside of this construction, will not have this exact factorised form and are currently being investigated. 

It is interesting to note that these four-dimensional orbifolds are complex but not K\"ahler. The holomorphic two-form, $\Omega$, is 
\be
\Omega=\bigg( \frac{x}{\sqrt{q(x)}}\dd x+\ii \frac{\sqrt{q(x)}}{2 x}\dd \psi\bigg)\wedge \bigg(\frac{y}{\sqrt{F(y)}}\dd y +\ii \frac{\sqrt{F(y)}}{\sqrt{h_1(y)h_2(y)}} Dz\bigg)\, ,
\ee
and one can check that
\be
\dd\Omega=\ii \bigg( -\frac{1}{4}\partial_x\Big(\frac{q(x)}{x^2}\Big)\dd\psi- \frac{\sqrt{h_1(y)h_2(y)}}{2y} \partial_y\Big(\frac{F(y)}{h_1(y)h_2(y)}\Big)Dz\bigg)\wedge \Omega=\ii P_{\rho} \wedge \Omega\, .
\ee
The putative compatible K\"ahler form, allowing for an arbitrary conformal factor $s(x,y)$, would then be
\be
J=s(x,y)\Big(\frac{y}{\sqrt{h_1(y)h_2(y)}}\dd y\wedge Dz+\frac{1}{x}\dd x\wedge \dd \psi\Big)\, .
\ee
It is not hard to convince one-self that this can never be closed for any choice of the conformal factor. One may wonder whether it is possible to make this K\"ahler for other uplifts. A necessary condition is that the gauge field strength, $\mathcal{F}$, is proportional to the volume form on the horizon of the four-dimensional solution by a constant. If this is the case, one can then solve for the conformal factor in order to obtain a K\"ahler space. This is, of course, not a generic condition to impose and is probably restrictive to the point that it is only possible for constant curvature Riemann surface horizons and, for the four-dimensional minimal theory, this means a hyperbolic horizon. For the constant curvature solutions that are known, \cite{Suh:2018tul,Hosseini:2018usu,Suh:2018szn}, the analogous four-dimensional spaces are K\"ahler. It would be interesting to classify the supersymmetric AdS$_2\times M_4$ spaces of six-dimensional U$(1)^2$-gauged supergravity in order to understand the underlying geometry of $M_4$ better. 

From the exterior derivative of the holomorphic two-form, $\Omega$, above we may extract the Ricci-form potential $P_{\rho}$, and by taking an exterior derivative the Ricci-form $\rho=\dd P_{\rho}$. Recall that for a complex manifold the Ricci-form $\rho$ is proportional to the first Chern class $c_1(M_4)$: $\rho=2\pi c_1(M_4)$. We may then compute the first Chern class threading through the various two-cycles using the Ricci-form. For the two-cycles $S_{\pm}$ we find
\be
\frac{1}{2\pi}\int_{S_{\pm}}\rho=\chi(\mathbbl{\Sigma}_1) \pm \frac{t}{ n_{\pm}m_+ m_-}\, ,
\ee
while for the spindle $\mathbbl{\Sigma}_2$ we have
\be
\frac{1}{2\pi}\int_{\mathbbl{\Sigma}_2}\rho=\chi(\mathbbl{\Sigma}_2)\,.
\ee
We may also compute the Euler characteristic of the four-dimensional orbifold. The Chern--Gauss-Bonnet theorem gives
\be
\chi(M_4)=\frac{1}{32 \pi^2}\int_{M_4} \text{vol}(M_4)\Big[ |\text{Riemann}|^2- 4 |\text{Ric}^{\text{tr}}|^2\Big]\,,
\ee 
where the first term is the contraction of the Riemann tensor into itself with no $4!$ weight, and the second term is the contraction of the traceless Ricci tensor into itself where
\be
\text{Ric}^{\text{tr}}_{\mu\nu}=R_{\mu\nu}-\frac{1}{4} R g_{\mu\nu}\,.
\ee
For the four-dimensional orbifold, after a tedious but otherwise simple computation, we find
\be
\chi(M_4)=\chi(\mathbbl{\Sigma}_1)\chi(\mathbbl{\Sigma}_2)\, .
\ee
Similarly we may compute the signature of the space, 
\be
\tau(M_4)= \frac{1}{3}\int_{M_4}p_1(M_4)\, ,
\ee
with $p_1(M_4)$ the first Pontryagin class of the four-manifold. Explicit computation gives
\be
\tau(M_4)=\frac{(n_+^2-n_-^2) t}{ 3m_-m_+n_-^2n_+^2}\, .
\ee
Note that the signature vanishes if we take the $n_{\pm}\rightarrow 1$ limit, in which case $\mathbbl{\Sigma}_2$ becomes a round two-sphere. For the Hirzebruch surface we have
\be
\chi=\chi(\mathbb{CP}^1_1)\chi(\mathbb{CP}^1_2)=4\, ,\qquad \tau= 0\, ,
\ee
and, therefore, in the (formal) limit where both spindles become $\mathbb{CP}^1$ we indeed obtain the correct invariants for the Hirzebruch surface. 

As a final comment let us understand the eight-dimensional compact spaces appearing in the uplifted solutions. This is a four-dimensional hemisphere bundle over $M_4$. We can understand the hemisphere as being embedded in $\mathbb{R}\oplus \mathbb{C}\oplus\mathbb{C}$ where the two gauge fields, $A_i$, each fibre one of the copies of $\mathbb{C}$. It is interesting to note that the total space of the $\mathbb{C}^2$ bundle over $M_4$ is not Calabi--Yau. To see this, note that the Calabi--Yau condition requires that the integrated twisting conditions of the R-symmetry vector, $F^R$, should cancel with the first Chern class of $M_4$ integrated through the same two-cycle. For the two cycle, $\mathbbl{\Sigma}_2$, these are equal, however, through the other two-cycles this is not the case. At the section, $S_{\pm}$, the tangent bundle splits into a direct sum where the complex tangent bundle to the sections is $\mathcal{O}(m_++m_-)$ with Chern number $\tfrac{m_++m_-}{m_+m_-}$. The normal bundle is $\mathcal{O}(-t)$ and has Chern number, $-\frac{t}{n_{\pm} m_+m_-}$. The normal direction has a $\mathbb{Z}_{n_{\pm}}$ singularity which gives rise to the additional $n_{\pm}$ in the Chern number. Instead we find that the integral of the R-symmetry gauge field has $m_++m_-$ replaced with $m_+-m_-$, see \eqref{eq:FRoverS}. This is a consequence of taking a spindle in the four-dimensional theory with the anti-twist, which turns out to be the only option in the minimal theory. A truncation to a more general theory, $i.e.$, the four-dimensional U$(1)^2$ theory, would allow us to uplift both twist and anti-twist solutions, \cite{Ferrero:2021etw,Couzens:2021cpk}, and thus obtain Calabi--Yau solutions in this framework, too.

\vspace{2.6cm}

%%%%%%%%%%%%%%%%%%%%%%%%%%%%%%%%%%%%%
\section{AdS$_2\times\Sigma_\mathfrak{g}\ltimes\mathbbl{\Sigma}_2$ solutions}\label{sec:spinriem}
%%%%%%%%%%%%%%%%%%%%%%%%%%%%%%%%%%%%%

In four-dimensional minimal gauged supergravity, one can also obtain supersymmetric AdS$_2\times\Sigma_\mathfrak{g}$ solutions which are the near-horizon limit of the black hole solutions constructed in \cite{Romans:1991nq, Caldarelli:1998hg} where $\Sigma_\mathfrak{g}$ is a Riemann surface of genus $\mathfrak{g}>1$. The metric and the gauge field are\footnote{This solution can actually be obtained by taking a scaling limit of the spindle solution discussed earlier. See appendix \ref{app:globalsols} where we take this limit in the six-dimensional case. The four-dimensional case works similarly.  }
\begin{equation}
ds_4^2\,=\,\frac{1}{4}ds_{\text{AdS}_2}^2+\frac{1}{2}ds_{\Sigma_\mathfrak{g}}^2\,, \qquad F_{(2)}\,=\,\text{vol}_{\Sigma_\mathfrak{g}}\,.
\end{equation}

By employing the consistent truncation ansatz in \eqref{upansatz1}, \eqref{upansatz2} and \eqref{upansatz3}, we can uplift the AdS$_2\times\Sigma_\mathfrak{g}$ solution to $F(4)$ gauged supergravity and obtain AdS$_2\times\Sigma_\mathfrak{g}\ltimes\mathbbl{\Sigma}_2$ solutions,
\begin{align}
ds_6^2\,=&\,\left(y^2h_1(y)h_2(y)\right)^{1/4}\left[\frac{1}{4}ds_{\text{AdS}_2}^2+\frac{1}{2}ds_{\Sigma_\mathfrak{g}}^2\right. \notag \\
& \,\,\,\,\,\,\,\,\,\,\,\,\,\,\,\,\,\,\,\,\,\,\,\,\,\,\,\,\,\,\,\,\,\,\,\,\,\,\,\,\,\,\,\,\,\,\,\,\,\,\,\,\,\,\,\,\, \left.+\frac{y^2}{F(y)}dy^2+\frac{F(y)}{h_1(y)h_2(y)}\left(dz-\frac{1}{2m}\omega\right)^2\right]\,, \notag \\
A_i\,=&\,-\frac{y^3}{h_i(y)}\left(dz-\frac{1}{2m}\omega\right)\,, \notag \\
B\,=&\,-\frac{1}{2m}y\,\text{vol}_{\text{AdS}_2}\,, \notag \\
X_i\,=&\,\left(y^2h_1(y)h_2(y)\right)^{3/8}h_i(y)^{-1}\,,
\end{align}
where $\omega$ is the Levi-Civita one-form satisfying $d\omega=-\text{vol}_{\Sigma_\mathfrak{g}}$ and the volume of the Riemann surface is $\int_{\Sigma_{\mathfrak{g}}}\text{vol}_{\Sigma_{\mathfrak{g}}}=4\pi\left(\mathfrak{g}-1\right)$.

We see that the spindle, $\mathbbl{\Sigma}_2$, is non-trivially fibered over the Riemann surface, $\Sigma_{\mathfrak{g}}$.
In order for this fibration to be well-defined, the one-form from the fibration, $\eta=\frac{2\pi}{\Delta{z}}\left(dz-\frac{1}{2m}\omega\right)$, should be globally defined,
\begin{equation}
\frac{1}{2\pi}\int_{\Sigma_{\mathfrak{g}}}d\eta\,=\,t \in\mathbb{Z}\,.
\end{equation}
However, since $d\eta=\frac{2\pi}{\Delta{z}}\frac{1}{2m}\text{vol}_{\Sigma_{\mathfrak{g}}}$, using \eqref{delz}, we find a condition on the spindle numbers, $n_\pm$, the flux numbers, $p_1$, $p_2$, and the genus, $\mathfrak{g}$,
\begin{equation}
t\,=\,\left(\mathfrak{g}-1\right)\frac{n_+n_-}{n_++n_-}\frac{16gx^2}{3m\left(x^2+3\right)\left(\mu-x\right)}\,\in\,\mathbb{Z}\,.
\end{equation}
This condition ensures that away from the poles on the $\mathbbl{\Sigma}_2$ fibre, the space, $\Sigma_{\mathfrak{g}}\ltimes\mathbbl{\Sigma}_2$, is well-defined, \cite{Cheung:2022ilc}. However, the orbifold singularities remain at the poles of $\mathbbl{\Sigma}_2$. 

Let us define two-cycles, $S_a\equiv\{y=y_a\}$, $a=\pm$, to be the section defined at the two poles of the fibre, $\mathbbl{\Sigma}_2$, \cite{Cheung:2022ilc}. The fluxes threading through the two-cycles have charges,
\begin{align}
\frac{1}{2\pi}\int_{S_+}F_i\,=&\,-\frac{1}{m}\frac{y_+^3}{y_+^3+q_i}\left(\mathfrak{g}-1\right)\,, \notag \\
\frac{1}{2\pi}\int_{S_-}F_i\,=&\,-\frac{1}{m}\frac{y_-^3}{y_-^3+q_i}\left(\mathfrak{g}-1\right)\,.
\end{align}
Recall that the R-symmetry gauge field flux is $F^R=g(F_1+F_2)$. Then, we have 
\begin{align}
\frac{1}{2\pi} \int_{S_+} F^R&= -2\left(\mathfrak{g}-1\right)+\frac{t}{n_+ }\, ,\nonumber \\
\frac{1}{2\pi} \int_{S_-} F^R&=-2\left(\mathfrak{g}-1\right)-\frac{t}{n_- }\, .\label{eq:FRoverSRiemann}
\end{align}
We may again use the homology relation, $S_+-S_-\,=\,t\,\mathbbl{\Sigma}_2$, \cite{Cheung:2022ilc}, which implies
\begin{equation}
\left(\frac{1}{2\pi}\int_{S_-}F_1+\frac{1}{2\pi}\int_{S_-}F_2\right)-\left(\frac{1}{2\pi}\int_{S_+}F_1+\frac{1}{2\pi}\int_{S_+}F_2\right)\,=\,t\left(\frac{1}{2\pi}\int_{\mathbbl{\Sigma}_2}F_R\right)\,.
\end{equation}
We find that the R-symmetry flux through the spindle, $\mathbbl{\Sigma}_2$, is 
\begin{equation}
\frac{1}{2\pi}\int_{\mathbbl{\Sigma}_2}F^R\,=\,\frac{1}{n_-}+\frac{1}{n_+}\,=\frac{1}{2\pi}\,\int_{\mathbbl{\Sigma}_2}c_1\left(\mathbbl{\Sigma}_2\right)\,,
\end{equation}
where $c_1$ is the first Chern class as before.

The four-form flux threading through the orbifold four-cycle, $\Sigma_\mathfrak{g}\ltimes\mathbbl{\Sigma}_2$, is
\begin{align}
\frac{1}{\left(2\pi{l}_s\right)^3}\int_{\Sigma_\mathfrak{g}\ltimes\mathbbl{\Sigma}_2}F_{(4)}\,&=\,\frac{\mathfrak{g}-1}{3\pi}\left[g^2\left(y_+^2-y_-^2\right)\right]\left[g\Delta{z}\right]N \nonumber\\
&=\left(\mathfrak{g}-1\right)\frac{n_+ +n_-}{n_- n_+}\frac{3 \mu(\tau^2+1)- \tau(\tau^2+5)}{4\tau (\tau^2+3)}N\,,
\end{align}
where we employed \eqref{yns}, \eqref{delz}, and \eqref{fluxquanc}. Note that the quantities, $\left[g^2\left(y_+^2-y_-^2\right)\right]$ and $\left[g\Delta{z}\right]$, are independent of the gauge coupling, $g$. As before, the exact quantisation condition that we should impose on this flux number is still to be determined. One can certainly tune the parameters to make it integer, however, this may be too restrictive.

Finally we can calculate the Bekenstein--Hawking entropy in the same way as in the previous section,
\begin{equation}
S_{\text{BH}}\,=\,\left(\mathfrak{g}-1\right)\mathcal{F}_{S^3\times{\mathbbl{\Sigma}_2}}=\frac{A_h}{2\pi}\mathcal{F}_{S^3\times{\mathbbl{\Sigma}_2}}\,,
\end{equation}
with $\mathcal{F}_{S^3\times{\mathbbl{\Sigma}_2}}$ is given in \eqref{s3sig2free} and $A_h= 2\pi(\mathfrak{g}-1)$ the area of the horizon of the four-dimensional black hole. 

In this case we may again study whether the internal eight-dimensional manifold is Calabi--Yau. As one may expect given our earlier discussion it is Calabi--Yau in this case. This is, of course, a consequence of the twisting on the Riemann surface being performed by the regular topological twist rather than the anti-twist in the spindle example of the previous section.

%%%%%%%%%%%%%%%%%%%%%%%%%%%%%%%%%%%%%
\section{Conclusion}\label{sec:conclude}
%%%%%%%%%%%%%%%%%%%%%%%%%%%%%%%%%%%%%

In this work we have constructed a consistent truncation of U$(1)^2$-gauged supergravity in six dimensions to four-dimensional minimal gauged supergravity. Given that the truncation depends only on the local data of the two-dimensional surface on which we compactify, our truncation can be used to reduce on any of the global completions: spindles, discs, domain walls, black bottles, Riemann surfaces and more.  By employing our consistent truncation we have constructed the AdS$_2\times\mathbbl{\Sigma}_1\ltimes\mathbbl{\Sigma}_2$ solutions of U$(1)^2$-gauged supergravity in six dimensions, where the compact space is a four-dimensional orbifold.\footnote{One can also replace $\mathbbl{\Sigma}_2$ by any of the solutions in appendix \ref{app:globalsols}.} In order to uplift our new solution to massive type IIA we have constructed the uplift of six-dimensional U$(1)^2$ gauged supergravity to massive type IIA on a four-dimensional hemisphere keeping all fields and allowing for independent Romans mass and gauge coupling. 

With the uplift to massive type IIA in hand we obtain AdS$_2\times\mathbbl{\Sigma}_1\ltimes\mathbbl{\Sigma}_2\ltimes\hat{S}^4$ solutions. These are natural candidates for the near-horizon geometries of six-dimensional black holes with a four-dimensional orbifold horizon. Alternatively one can view these as the holographic duals of 5d USp(2$N$) theory compactified on the four-dimensional orbifold. It is an interesting problem to identify more precisely these field theory duals, as has been performed for the M5-branes on a disc solutions in \cite{Bah:2021mzw,Bah:2021hei,Couzens:2022yjl,Bah:2022yjf}. As a first step we have computed the Bekenstein--Hawking entropy of the solutions with which we may compare to a field theory computation. 
As a further application of our uplift formula we have constructed AdS$_2\times\Sigma_{\mathfrak{g}}\ltimes\mathbbl{\Sigma}_2$ solutions and once again calculated their Bekenstein--Hawking entropy. Similar comments about identifying the dual field theories of these solutions are once again applicable here. In particular, it would be very interesting to calculate the Bekenstein--Hawking entropy for both classes of solutions using the AdS/CFT correspondence, \cite{Maldacena:1997re}, via a localization calculation in the dual field theory. For the black hole solutions with a near-horizon of the form, AdS$_2\times\Sigma_{\mathfrak{g}_1}\times\Sigma_{\mathfrak{g}_2}$, found in \cite{Suh:2018tul, Hosseini:2018usu, Suh:2018szn}, the Bekenstein--Hawking entropy was counted microscopically by the topologically twisted index of 5d USp(2$N$) theories, \cite{Hosseini:2018uzp, Crichigno:2018adf}. One expects that a modification of that computation should be possible here. 

We have focused on constructing AdS$_2$ solutions in this work which are natural candidates for the near-horizon geometries of six-dimensional black holes. It would be interesting to construct the full black hole solutions, which flow from an AdS$_6$ asymptotic regime to the near-horizon geometries studied here. One can then uplift this full black hole solution and study its properties directly in ten dimensions, see, for example, \cite{Suh:2019ily}, for the uplift of the AdS$_2\times\Sigma_{\mathfrak{g}_1}\times\Sigma_{\mathfrak{g}_2}$ solutions of \cite{Suh:2018tul}.  

A natural further extension of this work is to construct four-dimensional orbifolds where the parameter, $t$, appearing in \eqref{eq:tdef} is a free parameter. Using our construction this is not possible since the uplift formula relates this parameter to the magnetic charge threading through horizon of the four-dimensional solution and thus cannot be made an independent parameter, leading to the constraints we saw earlier. One must then construct these more general four-dimensional orbifolds directly in the six-dimensional theory. Work in this direction is currently ongoing and will be published soon. 

Finally, we note that it would be interesting to construct truncations which allow for a larger number of fields in the truncation, $i.e.$, beyond the minimal case considered here and in \cite{Cheung:2022ilc}. This would open up the possibility of constructing four-dimensional orbifolds of the type studied here where the spindle $\mathbbl{\Sigma}_1$ allows for both the twist and anti-twist solutions rather than just the anti-twist case. In four dimensions this would require being able to uplift the four-dimensional $T^3$ theory which allows both twist and anti-twist solutions. In \cite{Hosseini:2020wag} they constructed such a consistent truncation on $\Sigma_\mathfrak{g}$ for $\mathfrak{g}>1$. We argued earlier that our construction also gives the truncation on $\Sigma_\mathfrak{g}$ to four-dimensional minimal gauged supergravity after taking a particular scaling limit. It would be interesting to see if one can indeed construct a consistent truncation to the four-dimensional $T^3$ theory. 

%%%%%%%%%%%%%%%%%%%%%%%%%%%%%%%%%%%%%
\bigskip
\bigskip
\leftline{\bf Acknowledgements}
\noindent This research was supported by the National Research Foundation of
Korea under the grants, NRF-2022R1A2B5B02002247 (CC, HK, NK, YL), NRF-2020R1A2C1008497 (CC, HK, MS), and NRF-2018R1A2A3074631 (YL). Hospitality at APCTP during the program "Strings, Branes and Gauge theories 2022" is kindly acknowledged. In the latter stages of this work CC has been supported by the Mathematics Department of the University of Oxford.
%%%%%%%%%%%%%%%%%%%%%%%%%%%%%%%%%%%%%

%%%%%%%%%%%%%%%%%%%%%%%%%%%%%%%%%%%%%
\appendix
\section{The equations of motion}\label{app:EOMs}
%%%%%%%%%%%%%%%%%%%%%%%%%%%%%%%%%%%%%

For completeness we give the equations of motion of six-dimensional U$(1)^2$-gauged supergravity derived from the action in \eqref{actionaction}. The Einstein equation is
\begin{align}
&R_{\mu\nu}-\frac{1}{2}\sum_{i=1}^2\partial_\mu\varphi_i\partial_\nu\varphi_i-\frac{1}{4}Vg_{\mu\nu}-\frac{1}{2}\sum_{i=1}^2X_i^{-2}\left(F_{i\mu\rho}F_{i\nu}\,^\rho-\frac{1}{8}g_{\mu\nu}F_{i\rho\sigma}F_i\,^{\rho\sigma}\right) \nonumber \\
-&\frac{m^2}{4}\left(X_1X_2\right)^{-1}\left(B_{\mu\rho}B_\nu\,^\rho-\frac{1}{8}g_{\mu\nu}B_{\rho\sigma}B^{\rho\sigma}\right)-\frac{1}{16}\left(X_1X_2\right)^2\left(H_{\mu\rho\sigma}H_\nu\,^{\rho\sigma}-\frac{1}{6}g_{\mu\nu}H_{\rho\sigma\lambda}H^{\rho\sigma\lambda}\right)\,=\,0\,,
\end{align}
where the scalar fields, $\varphi_i$, can be read off from
\begin{equation}
X_i\,=\,e^{-\frac{1}{2}\vec{a}_i\cdot\vec{\varphi}}\,, \qquad \vec{a}_1\,=\,\left(2^{1/2},2^{-1/2}\right)\,, \qquad \vec{a}_2\,=\,\left(-2^{1/2},2^{-1/2}\right)\,.
\end{equation}
This is supplemented by the following equations of motion for the scalar fields and fluxes,
\begin{align}
\dd\star_6 \dd \log \left(\frac{X_1}{X_2}\right)&=-\frac{4 g m (X_1-X_2)}{X_1^{3/2} X_2^{3/2}}\star 1 -X_1^{-2}F_1\wedge \star_6 F_1+X_2^{-2} F_2\wedge \star_6 F_2\, , \nonumber \\
\dd \star_6 \dd \log (X_1X_2)&=-\frac{3 m^2 +4 g^2 X_1^4 X_2^4 -4 g m X_1^{3/2} X_2^{3/2} (X_1+X_2)}{2 X_1^3 X_2^3}\star_6 1\nonumber\\
&-\frac{1}{4}\Big(X_1^{-2} F_1\wedge \star_6 F_1 +X_2^{-2} F_2\wedge \star_6 F_2\Big) +\frac{X_1^2X_2^2}{8}H\wedge \star_6 H-\frac{m^2}{8 X_1 X_2}B\wedge \star_6 B\, ,
\end{align}
\begin{align}
\dd \Big(X_1^{-2} \star_6 F_1\Big)&=-\frac{1}{2} H\wedge F_2\, ,\nonumber\\
\dd \Big(X_2^{-2} \star_6 F_2\Big)&=-\frac{1}{2} H\wedge F_1\, ,\nonumber\\
\dd \Big(X_1^2 X_2^2 \star_6 H\Big)&=\frac{2 m^2}{X_1 X_2}\star_6 B+2 F_1\wedge F_2 +\frac{m^2}{2} B\wedge B\,.
\end{align}

The supersymmetry variations of the fermionic fields are\footnote{The Killing spinor equations of $F(4)$ gauged supergravity are traditionally written in terms of a pair of symplectic Majorana spinors $\eta^{A}$. Instead we will define a Dirac spinor from this pair by $\eta=\eta^1+\eta^2$ since this is most convenient for our later use.}
\begin{align}
\delta\psi_\mu\,=&\,\mathcal{D}_\mu\eta+\frac{1}{8}\left[g\left(X_1+X_2\right)+mX_0\right]\gamma_\mu\eta+\frac{i}{32}\left(X_1^{-1}F_1+X_2^{-1}F_2\right)_{\nu\lambda}\left(\gamma_\mu\,^{\nu\lambda}-6\delta_\mu^\nu\gamma^\lambda\right)\eta \notag \\
+&\frac{1}{32}m\left(X_1X_2\right)^{-1/2}B_{\nu\lambda}\left(\gamma_\mu\,^{\nu\lambda}-6\delta_\mu^\nu\gamma^\lambda\right)\gamma_7\eta+\frac{1}{96}X_1X_2H_{\nu\lambda\rho}\gamma^{\nu\lambda\rho}\gamma_7\gamma_\mu\eta\,, \label{eq:Grav6d}\\
\delta\chi\,=&\,\frac{1}{4}\partial_\mu\log\left(X_1X_2\right)\gamma^\mu\eta-\frac{1}{8}\left[g\left(X_1+X_2\right)-3mX_0\right]\eta-\frac{i}{32}\left(X_1^{-1}F_1+X_2^{-1}F_2\right)_{\mu\nu}\gamma^{\mu\nu}\eta \notag \\
-&\frac{1}{32}m\left(X_1X_2\right)^{-1/2}B_{\mu\nu}\gamma^{\mu\nu}\gamma_7\eta+\frac{1}{96}X_1X_2H_{\mu\nu\lambda}\gamma^{\mu\nu\lambda}\gamma_7\eta\,, \\
\delta\lambda\,=&\,\frac{1}{2}\partial_\mu\log\left(\frac{X_1}{X_2}\right)\gamma^\mu\eta-g\left(X_1-X_2\right)\eta-\frac{i}{4}\left(X_1^{-1}F_1-X_2^{-1}F_2\right)_{\mu\nu}\gamma^{\mu\nu}\eta\,,
\end{align}
where
\begin{equation}
\mathcal{D}_\mu\eta=\nabla_\mu\eta-\frac{i}{2}g\left(A_1+A_2\right)_\mu\eta\,.
\end{equation}

%%%%%%%%%%%%%%%%%%%%%%%%%%%%%%%%%%%%%
\section{The complete uplift formula to massive type IIA }\label{app:Uplift}
%%%%%%%%%%%%%%%%%%%%%%%%%%%%%%%%%%%%%

For the vanishing two-form field, $B=0$, the uplift formula of U$(1)^2$-gauged supergravity in six dimensions was first presented in \cite{Cvetic:1999xx} and improved in \cite{Faedo:2021nub}. In this appendix we present the complete uplift formula with non-trivial two-form field, $B$, and no requirement on a relationship between the Romans mass and the gauge coupling. This will allow for the uplift of non-supersymmetric solutions on the branch for which $3m\neq 2g$, for example, which were not possible to uplift previously using the results of \cite{Faedo:2021nub}. We follow the conventions in \cite{Faedo:2021nub} so that the truncation formula we present recover those of \cite{Faedo:2021nub} upon setting $B=0$ and $3m=2g$.

The bosonic field content of massive type IIA supergravity, \cite{Romans:1985tz}, consists of the metric, the dilaton field, $\Phi$, the NSNS two-form potential, $B_{(2)}$, the RR one- and three-form potentials, $C_{(1)}$ and $C_{(3)}$, and the Romans mass, $F_{(0)}$. The field strengths of the potentials are 
\begin{equation}\label{eq:IIApotentials}
H_{(3)}\,=\,dB_{(2)}\,, \quad F_{(2)}\,=\,dC_{(1)}+F_{(0)}B_{(2)}\,, \quad F_{(4)}\,=\,dC_{(3)}+B_{(2)}\wedge F_{(2)}-\frac{1}{2}F_{(0)} B_{(2)}\wedge B_{(2)}\,.
\end{equation}
The action in string frame is 
\begin{align}
S\,&=\,\frac{1}{16\pi{G}_N^{(10)}}\left\{\int{d}^{10}x\sqrt{-g}\left[e^{-2\Phi}\left(R+4|d\Phi|^2-\frac{1}{2}|H_{(3)}|^2\right)-\frac{1}{2}\left(F_{(0)}^2+|F_{(2)}|^2+|F_{(4)}|^2\right)\right]\right. \notag \\
&-\frac{1}{2}\int\left(B_{(2)}\wedge{d}C_{(3)}\wedge{d}C_{(3)}+\frac{1}{3}F_{(0)}B_{(2)}\wedge{B}_{(2)}\wedge{B}_{(2)}\wedge{d}C_{(3)}\right. \notag \\
& \qquad\qquad\qquad\qquad\qquad\qquad\qquad\qquad \left.\left.+\frac{1}{20}F_{(0)}^2B_{(2)}\wedge{B}_{(2)}\wedge{B}_{(2)}\wedge{B}_{(2)}\wedge{B}_{(2)}\right)\right\}\,.
\end{align}
At the level of the classical equations of motion, there are two scaling symmetries of the fields which preserve the equations of motion. Let $\lambda$ and $l$ denote the two constant scales. Then the equations of motion are symmetric under
\begin{align}
d&\hat{s}_{\text{s.f.}}^2\,=\,\lambda^2 l^2 ds_{\text{s.f.}}^2\,, \qquad e^{\hat{\Phi}}\,=\,\lambda^2e^\Phi\,, \qquad \hat{B}_{(2)}\,=\,\lambda^2l^2 B_{(2)}\,, \notag \\
&\hat{F}_{(0)}\,=\,\lambda^{-3}l^{-1}F_{(0)}\,, \qquad \hat{C}_{(n-1)}\,=\,\lambda^{n-3}l^{n-1}C_{(n-1)}\,,
\end{align}
with $n=2,4$. The scaling parameter, $\lambda$, plays an important role for the fluxes to be properly quantized, however, once the flux quantisation has been imposed, the scaling symmetry is broken, \cite{Faedo:2021nub}. On the other hand, the parameter, $l$, is a length scale one can introduce but could equally reabsorb it into the string length in the quantisation condition.

The complete uplift formula of U$(1)^2$-gauged supergravity in six dimensions to massive type IIA supergravity with arbitrary Romans mass and gauge coupling, has string frame metric,
\begin{align}
\dd s^2_{\text{s.f.}}\,&=\,\lambda^2l^2\mu_0^{-1/3}\left(X_1X_2\right)^{-1/4}\Delta^{1/2}\left\{\dd s_6^2\right. \notag \\
&\left.+g^{-2}\Delta^{-1}\left[X_0^{-1}d\mu_0^2+X_1^{-1}\left(\dd\mu_1^2+\mu_1^2D\phi_1^2\right)+X_2^{-1}\left(\dd\mu_2^2+\mu_2^2D\phi_2^2\right)\right]\right\}\,, \\
e^\Phi\,&=\,\lambda^2\mu_0^{-5/6}\Delta^{1/4}\left(X_1X_2\right)^{-5/8}\,,
\end{align}
where the function, $\Delta$, is
\begin{equation}
\Delta\,=\,\sum_{a=0}^2X_a\mu_a^2\,,
\end{equation}
and the one-forms are $D\phi_i=d\phi_i-gA_i$. The angular coordinates, $\phi_i$, have canonical $2\pi$ periodicity. The $\mu$'s are embedding coordinates satisfying  $\sum_{a=0}^2\mu_a^2=1$ and a convenient parametrisation to employ is
\begin{equation}
\mu_0\,=\,\cos\xi\,, \qquad \mu_1\,=\,\sin\xi\sin\eta\,, \qquad \mu_2\,=\,\sin\xi\cos\eta\,,
\end{equation}
where $\eta\in[0,\pi/2]$, $\xi\in[0,\pi/2]$. The restricted range of the $\xi$ coordinate in comparison with the usual range on an $S^4$ is a result of considering the four-hemisphere as opposed to a round four-sphere, this is required due to the overall $\mu_0$ factor of the metric. The four-form flux is given by
\begin{align} 
\lambda^{-1}l^{-3}F_{(4)}\,&=\sqrt{\frac{2g}{3m}}\frac{\mu_0^{-2/3}}{2 g^3 \Delta}\bigg[ \frac{U}{2 \Delta}\dd\mu_1^{2}\wedge\dd\mu_2^2\wedge D\phi_1\wedge D\phi_2\nonumber\\
&-\frac{X_1 X_2}{\Delta}\bigg(\frac{\mu_0^2 X_0^2}{X_1X_2}\Big(\mu_1^2\dd \frac{X_1}{X_0}\wedge \dd\mu_2^2-\mu_2^2\dd\frac{X_2}{X_0}\wedge \dd\mu_1^2\Big)-\mu_1^2\mu_2^2\dd\log\frac{X_1}{X_2}\wedge\dd\mu_0^2\bigg)\wedge D\phi_1\wedge D\phi_2\nonumber\\
&+g\Big( F_1\wedge \big(X_2 \mu_2^2\dd\mu_1^2+(X_0 \mu_0^2 +X_2 \mu_2^2)\dd\mu_2^2\big)\wedge D\phi_2\nonumber\\
&\qquad\qquad+F_2\wedge \big(X_1 \mu_1^2\dd\mu_2^2+(X_0 \mu_0^2+X_1\mu_1^2)\dd\mu_1^2\big)\wedge D\phi_1\Big)\bigg]\nonumber\\
&+\sqrt{ \frac{2g}{3m}}\bigg[ \frac{1}{4g}X_1^2 X_2^2 \mu_0^{-2/3} \star_6 H \wedge \dd\mu_0^2-\frac{3 m^2}{4g} \frac{\mu_0^{4/3}}{X_1 X_2}\star_6 B\bigg]\, ,\label{eq:F4exp}
\end{align}
where $\star_6$ is the Hodge dual with respect to the six-dimensional metric, $ds_6^2$, and the function, $U$, is defined to be
\begin{equation}
U\,=\,2\sum_{a=0}^2X_a^2\mu_a^2-\left[\frac{4}{3}X_0+2\left(X_1+X_2\right)\right]\Delta\,.
\end{equation}
From the above, one finds the following three-form potential for $F_{(4)}$, as defined in \eqref{eq:IIApotentials},
\begin{align}
\lambda^{-1}l^{-3} C_{(3)}&=\frac{1}{2g^3}\sqrt{\frac{2g}{3m}}\bigg[ \mu_0^{-2/3}\Big(\frac{1}{2}\big(\dd\mu_2^2-\dd\mu_1^2\big)+\frac{1}{\Delta}(X_2 \mu_2^2\dd\mu_1^2-X_1 \mu_1^2\dd\mu_2^2)\Big)\wedge D\phi_1\wedge D\phi_2\nonumber\\
&-\frac{3 g \mu_0^{4/3}}{4}(F_1\wedge D\phi_2+F_2\wedge D\phi_1)-\frac{3 g^2}{4}\mu_0^{4/3}X_1^2X_2^2 \star_6 H\bigg]\, .
\end{align}
For completeness, the Hodge dual of the four-form, $F_{(4)}$, is
\begin{align}
\lambda^{-3} l^{-5}\star_{10}F_{(4)}&=g U  \text{vol}_6 -\frac{1}{2g^2}\sum_{i=1}^{2}X_i^{-2}\star_6 F_i\wedge \dd\mu_i^2\wedge D\phi_i-\frac{1}{2g}\sum_{a=0}^{2}X_{a}^{-1}\star_6\dd X_a\wedge \dd\mu_a^2\nonumber\\
&-\frac{1}{4g^3 \Delta}H\wedge (X_2 \mu_2^2\dd\mu_1^2-X_1\mu_1^2\dd\mu_2^2)\wedge D\phi_1\wedge D\phi_2\nonumber\\
&+\frac{X_0}{12 g^3 \Delta}B\wedge \dd\mu_1^2\wedge \dd \mu_2^2\wedge D\phi_1\wedge D\phi_2\, .
\end{align}
The two-form flux and the NSNS two-form potential are given, respectively, by
\begin{align}
\lambda l^{-1}{F}_{(2)}\,=&\,-\frac{m}{2}\mu_0^{2/3}B\,, \notag \\
\lambda^{-2}l^{-2}B_{(2)}\,=&\,-\frac{1}{2}\sqrt{\frac{3m}{2g}}\mu_0^{2/3}B\,.
\end{align}
Note that the RR one-form potential, $C_{(1)}$, is pure gauge. Finally the Romans mass is given by
\begin{equation}
\lambda^3 l F_{(0)}\,=\,\sqrt{\frac{2 g m}{3}}\,.
\end{equation}
For pure $F(4)$ gauged supergravity there are five inequivalent theories depending on the values of gauge coupling, $g$, and mass parameter, $m$. We have presented the uplift adapted for the supersymmetric fixed point where $3m=2g$ since this is the focus in the main text. However, for one of the other theories where either $g$ or $m$ vanish, or becoming negative, one should use the scaling symmetries so that the metric and form fields are smooth in the vanishing limit. 

%%%%%%%%%%%%%%%%%%%%%%%%%%%%%%%%%%%%%
\section{A cornucopia of global solutions}\label{app:globalsols}
%%%%%%%%%%%%%%%%%%%%%%%%%%%%%%%%%%%%%

In this appendix we will study in more detail the variety of solutions arising from different global completions of the local solution given in section \ref{sec:M2sols}. This complements the analysis of the spindle solution in the main text a similar analysis for the full zoo of possible solutions. 
Recall that the different global completions are characterised by the roots of the function $F(y)$. From figure \ref{fig:psdomain} we see that this is broken up into three regions, with special lines given by the various boundaries of these regions. In addition to previously studied solutions we also introduce new solutions which have not appeared previously in the literature. We find that there are solutions for arbitrary constant curvature Riemann surface, discs, black bottles, black goblets and domain wall solutions. 

Throughout this section we will be setting $3m=2g$ since it makes the analysis simpler, though it is not necessary unless one wishes to obtain a supersymmetric solution in ten dimensions.

%%%%%%%%%%%%%%%%%%%%%%%%%%%%%%%%%%%%%
\subsection{Discs}
%%%%%%%%%%%%%%%%%%%%%%%%%%%%%%%%%%%%%

The first solution we will study in this appendix are the disc solutions. There are two types to consider depending on how the space is ended. In both cases one end-point is a single root and the other point is at $0$. The difference between the two cases arises when considering whether the 0 is a root of $F(y)$ or not. For 0 to be a root of $F(y)$ it requires one of the $q_i$ to be set to zero, without loss of generality we take $q_2=0$. As we will see shortly, when considering this case, the uplifted metric in ten dimensions has an enhancement of symmetry, U$(1)\rightarrow$SU$(2)$. The second case, which does not have this symmetry enhancement, has $y\in [0,y_-]$ or $y\in[y_-,0]$ with $y_-$ a positive/negative root of $F(y)$ and $\pm F'(y_-)<0$. This type of disc has already appeared in the literature in \cite{Suh:2021aik}, albeit with a different choice of coordinates. We will first study the disc with enhanced symmetry in ten dimensions which has not appeared previously, before turning our attention to this second type of disc which has appeared in \cite{Suh:2021aik}. 

\paragraph{Disc with enhanced symmetry} We want $ F(y)$ to have a root at $0$ which is achieved without loss of generality by imposing $q_2=0$. In figure \ref{fig:psdomain} this corresponds to the line $p=0$. The function $F(y)$ simplifies to
\begin{equation}
m^4F\Big(\frac{y}{m}\Big)=y^3\bigg(y^3-y+Q_1\bigg)\equiv y^3 f(y)\, ,
\end{equation}
where $f(y)$ is now a cubic and we have defined $Q_1=m^3 q_1$. We can now fix $y$ to start at $y=0$. For the metric to be of the correct signature, and for the scalar fields to be non-negative, we require that the cubic $f(y)$ admits three real roots, two of which are positive. Then, $y\in[0,y_+]$ where $y_+$ is the smaller of the two positive roots. This requires us to fix 
\begin{equation}
0\leq Q_1\leq \frac{2}{3\sqrt{3}}\, .
\end{equation}
In figure \ref{fig:psdomain} this corresponds to the red line stretching between the origin and the point $B$.

At $y=y_+$ the metric looks locally like $\mathbb{R}^2/\mathbb{Z}_k$ if we fix the period of the $z$ coordinate to be 
\begin{equation}
\frac{\Delta z}{2\pi}=\frac{2}{m |f'(y_+)|k}\, .
\end{equation}
The degeneration of the metric at $y=0$ is singular, taking the form (we return to using the $q_1$ parameter rather than the $Q_1$)
\begin{equation}
\dd s^2_6=q_1^{1/4} r^{5/2}\bigg[\dd s^2_{\text{AdS}_4}+\frac{4}{m^2 q_1} \dd r^2+m^2 \dd z^2\bigg]\, ,
\end{equation}
where we have introduced the coordinate $r$ via $y=r^2$. It is not hard to convince oneself that this metric is singular at $r=0$ given the overall factor of $r$. The metric is conformal to the product metric on AdS$_4\times I\times S^1$, with $I$ a line interval. In addition to a singular metric the scalar fields are also singular at $y=r^2=0$, observe that near $y=0$ we have
\begin{equation}
X_1\sim q_1^{-5/8} r^{15/4}\, ,\qquad X_2 =X_0 \sim q_1^{3/8} r^{-9/4}x
\, .
\end{equation}

Similarly to the spindle in the main text we may compute the Euler characteristic of the disc. One must be slightly more careful since the disc has a boundary which can in principle contribute to the Euler characteristic, however since the boundary has vanishing geodesic curvature there is no such contribution here. After the dust settles one finds 
\begin{equation}
\chi(\mathbb{D})=\frac{1}{k}\, .
\end{equation}
Having set $q_2=0$ there is a solitary magnetic charge coming from the gauge field $A_1$. The magnetic charge is
\begin{equation}
\frac{g}{2\pi}\int_{\mathbb{D}}F_1=-\frac{g \Delta z}{2\pi}\frac{y_+^3}{h_1(y_+)}=\frac{1}{k}-\frac{m \Delta z}{4\pi}\, .
\end{equation}
We see that once again supersymmetry is not preserved by a topological twist but rather a different mechanism entirely. This twist is common to all discs found in the literature where symmetry is enhanced when comparing to the symmetry of the uplifted generic local solution. It would be interesting to understand if this is the only twist possible or whether there are more possibilities that have so far been missed. For spindles it was shown that only the twist and anti-twist are possible in \cite{Ferrero:2021etw}, it would be interesting to perform a similar analysis for discs in this spirit. 

Since the solution has no $B$ field we may uplift this solution to massive type IIA using the uplift formula presented in \cite{Faedo:2021nub}, see appendix \ref{app:Uplift} where we extended their work to include the $B$ field.  It is useful to introduce explicit coordinates for the embedding coordinates $\mu_a$. Contrary to the choice used elsewhere in this paper, it is more convenient to use the following parametrisation
\begin{align}
\mu_0=\sqrt{1-\mu^2} \cos\theta\, ,\quad  \mu_2=\sqrt{1-\mu^2}\sin\theta\, ,\quad \mu_1=\mu\, ,
\end{align}
with $\mu\in[0,1]$ and $\theta\in [0,\pi/2)$.

The uplifted metric is
\begin{align}
\lambda^{-2}l^{-2} \dd s^2_{10}&= \mu_0^{-1/3} \tilde{\Delta}^{1/2} \sqrt{y}\bigg[ \dd s^2(\text{AdS}_4)+\frac{1}{yf(y)}\dd y^2+\frac{f(y)}{h_1(y)}\dd z^2\nonumber\\
&+\frac{1}{g^2 y^2 \tilde{\Delta}}\Big((1-\mu^2) y^3 (\dd\theta^2+\sin^2\theta \dd\phi_2^2)+h_1(y) (\dd\mu^2+\mu^2D\phi_1^2)\Big)\bigg]\, ,
\end{align}
where 
\begin{align}
\tilde{\Delta}=h_1(y)(1-\mu^2)+\mu^2y^3 \, ,\qquad f(y)=m^2 h_1(y)-y\, .
\end{align}

We now want to investigate the singularity at $y=0$. It is useful to define
\be
y=r^{2/3}\,,
\ee
and then expand the metric around $r=0$, giving
\begin{align}
\lambda^{-2}l^{-2}\dd s^2_{10}&=\mu_0^{-1/3}h_1(0)^{1/2}(1-\mu^2)^{1/2} r^{1/3}\bigg[\dd s^2(\text{AdS}_4) +m^2 \dd z^2\nonumber\\
&+\frac{1}{r^{4/3}g^2 h_1(0)}\Big( \dd r^2+ r^2 \big(\dd\theta^2+\sin^2\theta \dd\phi_2^2\big)+  \frac{h_1(0)}{1-\mu^2}(\dd \mu^2+\mu^2\dd\phi_1^2)\Big)\bigg]\, .
\end{align}
This is the near-horizon of the metric of a D4-brane inside the worldvolume of a D8-brane smeared along two directions, see for example \cite{Bah:2017wxp}. The dilaton also has the correct $r^{-\tfrac{1}{2}}$ scaling in this limit. There is a two-dimensional hemisphere parametrised by the $\theta,\phi_2$ coordinates, in the uplifted metric which is part of the SU$(2)_R\times $U$(1)_r$ symmetry group of the solution. This SCFT should arise from wrapping the 5d $\mathcal{N}=1$ SCFT with gauge group USp$(2N)$ on a twice punctured sphere, one regular and one irregular. The full details of such a field theory are lacking.  

\begin{figure}[t!]
\centering
\includegraphics[width=0.5\linewidth]{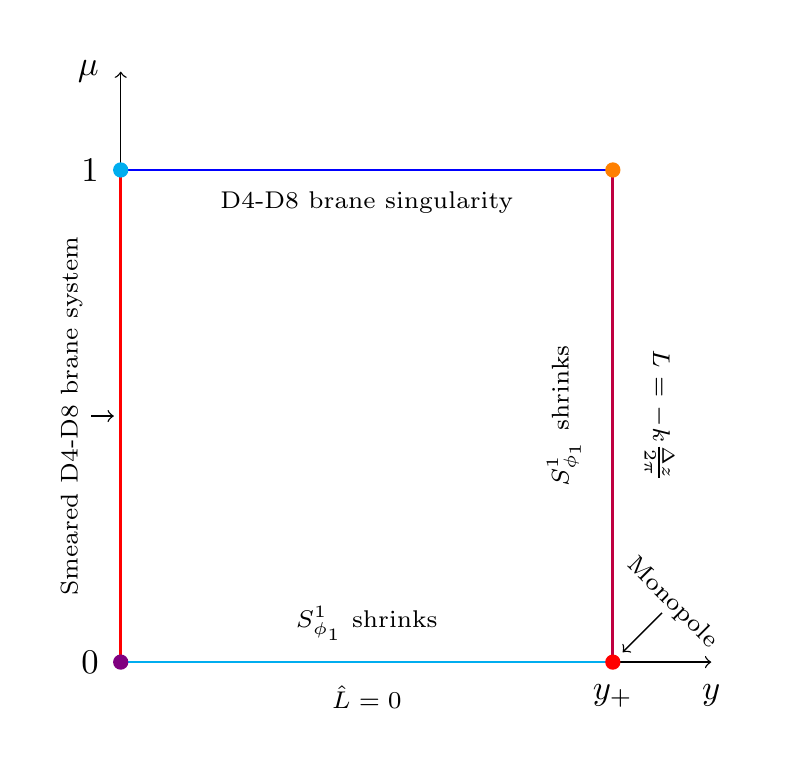}
  
%\captionsetup{width=0.75\linewidth}
\caption{A schematic plot of the rectangle over which the $S^2\times S^1_z\times S^1_{\phi_1}$ fibers are defined. Within the interior of the rectangle all the fibers remain of finite size. Along the boundary various cycles shrink. The red dot on the bottom right hand corner is the location of a monopole of charge $k$. }
\label{fig:rectangleD4N=2}
\end{figure}

To complete the regularity analysis of the solution one should change to a different set of coordinates which exhibits the metric as an $S^2\times S^1\times S^1$ fibration over a rectangle in a similar manner to that done in previous papers \cite{Couzens:2021rlk,Bah:2021mzw,Bah:2021hei}. It is also convenient to introduce a constant gauge shift $\delta A_1=\tfrac{1}{3} \dd z$. We find that the metric can be written in the form
\begin{align}
\lambda^{-2}l^{-2}\dd s^2&=\mu_0^{-1/3}\tilde{\Delta}^{1/2}\sqrt{y}\bigg[\dd s^2 (\text{AdS}_4)+\frac{y(1-\mu^2)}{g^2 \tilde{\Delta}}(\dd \theta^2+\sin^2\theta \dd\phi_2^2)\nonumber\\
&+\frac{1}{y f(y)}\dd y^2 +\frac{h_1(y)}{g^2 y^2\tilde{\Delta}}\dd\mu^2
+R_z^2 (\dd z+L \dd\phi_1)^2+ R_1^2\dd\phi_1^2\bigg]\, ,
\end{align}
where
\begin{align}
R_z^2&=\frac{S(y,\mu)}{y^2\big((1-\mu^2)h_1(y)+\mu^2 y^3\big)}\, ,\qquad R_1^2= \frac{\mu^2 f(y)}{g^2S(y,\mu)}\, ,\quad
 L=\frac{ \tfrac{1}{3}\mu^2(2 y^3-q_1)}{g S(y,\mu)}\, , \nonumber\\
 S(y,\mu)&=y^2\big((1-\mu^2)f(y)+m^2\mu^2y^3\big)+\tfrac{1}{3} \mu^2(\tfrac{1}{3} h_1(y)-2 y^3)\, .
\end{align}
We see that $R_z$ vanishes only at the point $y=y_+$, $\mu=0$, while $R_1$ vanishes at both $\mu=0$ and $y=y_+$ independently of the other. Moreover, the fibration term behaves as
\begin{align}
L(y,\mu=0)=0\, ,\qquad L(y=y_+,\mu)=-\frac{\Delta z}{2\pi} k\, .
\end{align}
This behaviour of the fibration signifies the presence of a monopole located at $(y=y_+,\mu=0)$. At $\mu=0$ the $\phi_1$ circle shrinks smoothly provided $\phi_1$ has period $2\pi$. At $\mu=1$ the metric degenerates as one would expect for a D4-brane inside a D8-brane. We have summarised the various degenerating cycles in figure \ref{fig:rectangleD4N=2}. 

It remains to consider the quantisation of the fluxes and to compute the free energy, however this is beyond the scope of this work.

\paragraph{Disc with reduced symmetry} Let us consider the second type of disc which was previouly studied in \cite{Suh:2021aik}. We require that $F(y)$ has a single root at $y_*$ and is positive in the domain $y\in[0,y_*]$. As before there is a conical deficit, with deficit angle $2\pi(1-k^{-1})$, at the single root, when $z$ has period
\be
\frac{\Delta z}{2\pi}=\frac{2 y_*^3}{mk |F'(y_*)|}\, .
\ee
At $y=0$ the metric is like in the disc above, singular, albeit in a different manner now. One can compute the Euler-characteristic of the solution, from \cite{Faedo:2021nub} or explicit computation we have
\be
 R\text{vol}(M_2)= \frac{\dd}{\dd y} \bigg(\frac{ F(y)\partial_y \big(h_1(y)h_2(y)\big)-h_1(y)h_2(y)\partial_y F(y)}{y \big(h_1(y)h_2(y)\big)^{3/2}}\bigg)\dd y \wedge \dd z\, .
\ee
The Euler characteristic is then given by
\be
\chi(\mathbb{D})=\frac{1}{4\pi} \int_{\mathbb{D}} R \text{vol}(\mathbb{D})=\frac{1}{k}\, .
\ee
Since the singular boundary part of the metric has vanishing geodesic curvature there are no additional contributions to the Euler characteristic from this boundary. The magnetic flux of the R-symmetry vector threading through the disc is
\be 
\frac{g}{2\pi}\int_{\mathbb{D}} (F_1+F_2)=\frac{1}{k} -\frac{m}{\pi} \Delta z\, .
\ee
This is of a similar flavour to the twist of the disc studied above, in both cases the twist is neither the usual topological twist nor the twist and anti-twist of the spindle. 

To analyse the singularity of the metric it is once again convenient to uplift the solution to ten dimensions. Expanding around $y=0$ the metric takes the form
\be
\dd s^2=\mu_0^{-1/3}\bigg[ \frac{\sqrt{q_1 q_2}\mu_0}{y}\Big( \dd s^2(\text{AdS}_4)+m^2 \dd z^2\Big)+\frac{y\mu_0}{m^2\sqrt{q_1 q_2}} \dd y^2+\frac{1}{g^2\sqrt{q_1 q_2}\mu_0}\sum_{i=1}^{2} q_i \big(\dd \mu_i^2+\mu_i^2 D\phi_i^2\big)\bigg]\, ,
\ee
which we identify as the metric on a smeared D4-D8 brane bound state, with the smearing over four transverse directions, \cite{Bah:2017wxp}.\footnote{This also allows for an interpretation as an O4-O8 bound state, see \cite{Passias:2018zlm}.} Contrast this with the previous disc where the smearing was only over two of the four directions. 
Note that after a suitable change of coordinates it is not hard to see that the four terms in the summation of the last part of the metric is actually flat. This degeneration is then substantially different to the more canonical disc with enhanced symmetry above and studied in the literature. This type of disc has been studied in \cite{Suh:2021aik} in pure $F(4)$ gauged supergravity and for M5-branes in \cite{Karndumri:2022wpu}, similar discs with this kind of degeneration are not possible where other discs have been found. It would be interesting to try to understand the field theory of these types of discs. The $\mathcal{N}=1$ version from wrapping M5-branes may be the most accessible given the advancements in understanding the field theory there of the $\mathcal{N}=2$ discs. From the form of the metric these require a `bare' $y$ coordinate which is not present in any of the lower dimensional gauged supergravity solutions known, only in six and seven dimensions. It is possible that this is an artefact of the theories previously considered rather than the dimension and it would be interesting to see if one can generate these types of discs in the lower dimensional gauged supergravities. 

One could now attempt to draw a similar rectangular diagram over which the isometries of the space are fibered. In this case the rectangle becomes three-dimensional cube and can be parametrised by $y,\mu_1,\mu_2$. We refrain from presenting this analysis further details can be found in \cite{Suh:2021aik}. 

%%%%%%%%%%%%%%%%%%%%%%%%%%%%%%%%%%%%%
\subsection{Black bottles and black goblets}
%%%%%%%%%%%%%%%%%%%%%%%%%%%%%%%%%%%%%

\begin{figure}[t!]
\centering
\includegraphics[width=0.38\linewidth]{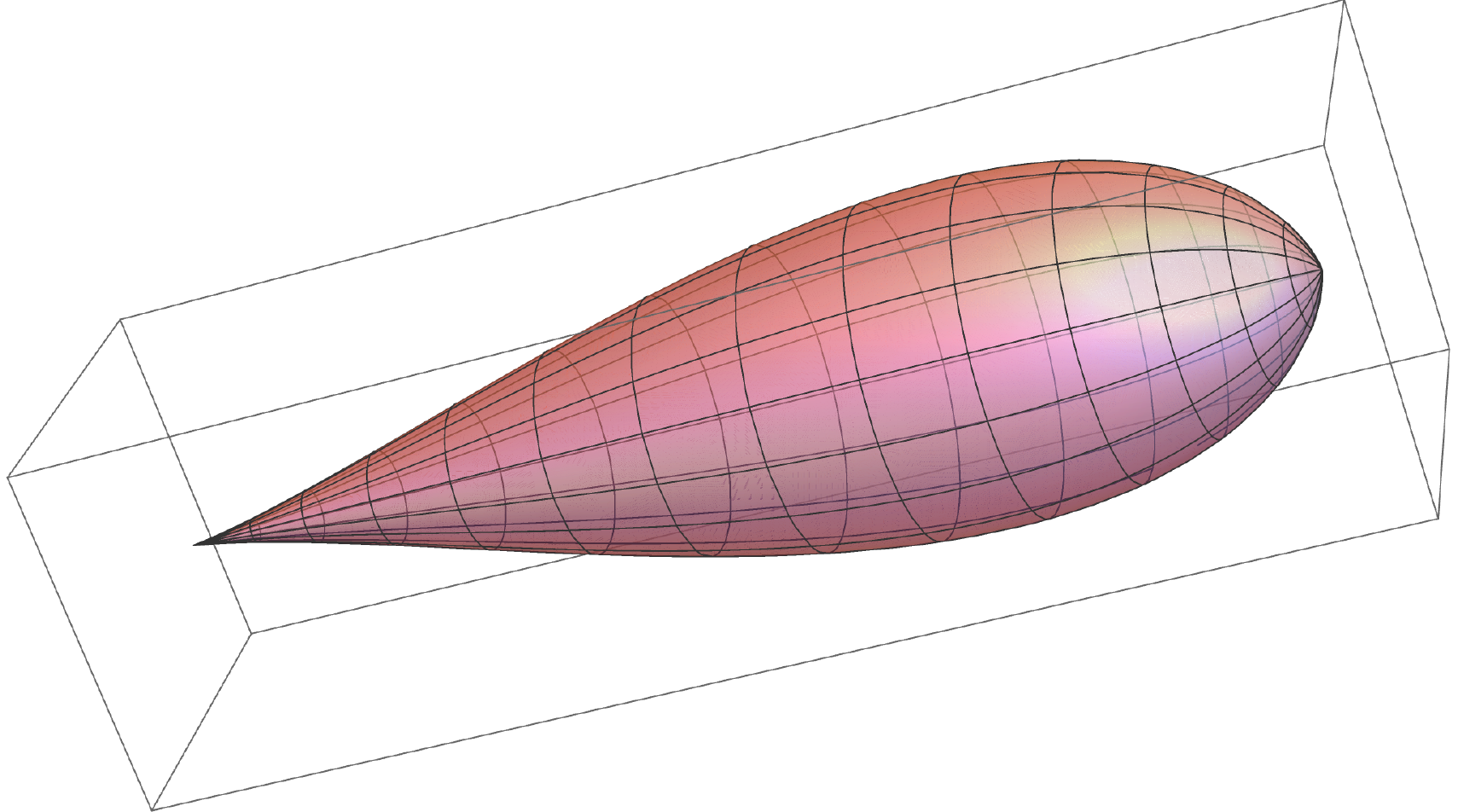} \includegraphics[width=0.38\linewidth]{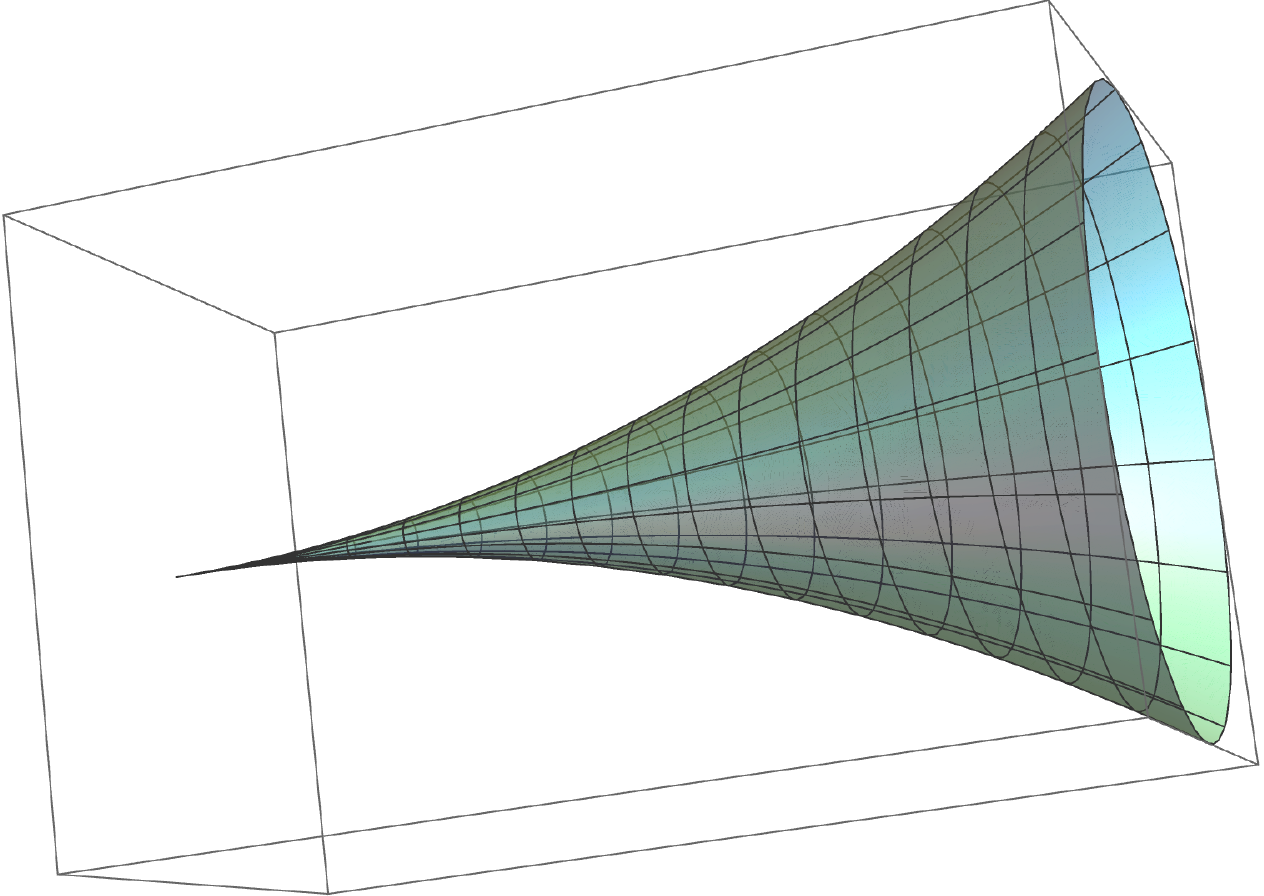}
%\captionsetup{width=0.75\linewidth}
\caption{Cartoons of a black bottle (left) and a black goblet (right). For the black bottle, the right end-point is a $\mathbb{R}^2/\mathbb{Z}_k$ singularity. The left end-point is cusp metric: this pole is infinitely far from the other pole but the volume is finite. For the black goblet, the right side is a circle and the left end-point is cusp metric.}
\label{fig:blackbottle}
\end{figure}

\paragraph{Black bottles} So far we have considered roots which are either single roots or zero, but what happens when we have a double root? For the Black bottle case the double root is the bigger of the two positive roots. Let us call the double root of $F(y)$ $y=y_+$, and write $F(y)=(y-y_+)^2 G(y)$ with $G(y_+)\neq 0$. Note that neither of the $h_i(y)$ vanish at this root. The metric takes the form
\be
\dd s^2_{6}=\frac{y_+^{3/2}}{\sqrt{m}}\bigg[\dd s^2_{\text{AdS}_4}+\frac{y_+^2}{G(y_+)}\Big( \dd\rho^2 +\me^{-2\rho}\dd \hat{z}^2\Big) \bigg]\, ,
\ee
where we used the change of coordinates $|y-y_+|=\me^{-\rho}$, $z=\frac{y_+^3}{G(y_+)m}\hat{z}$. The metric in the brackets is the metric of a cusp, it is non-compact but has finite area. For a smooth cusp the period of $\hat{z}$ is $2\pi$ however we may allow for deficit angles here too. We can now glue this double root with a single root ($y=y_-$) at the other end-point, thereby obtaining a conical singularity at that point. The resultant metric is non-compact but has finite area as we will see shortly. We have drawn a representative of a black bottle in the figure \ref{fig:blackbottle}. This type of double zero degeneration has appeared previously in the hep-th literature, see \cite{Klemm:2014rda,Hennigar:2015cja,Gnecchi:2013mja}, where it was considered at both end-points.\footnote{In the latter reference they obtained such solutions by taking an ultraspinning limit of the four-dimensional Kerr--Newman-AdS solution.} Black bottles have appeared in the gr literature, see for example \cite{Chen:2016rjt,Chen:2015zoa}. 

It would be interesting to compute the Euler characteristic in this case. One expects a contribution of $\frac{1}{n_+}$ from the regular conical singularity. The contribution from the cusp is more subtle. We will leave understanding  this to the future.

\paragraph{Black goblets} We can also generalise the black bottle to find a black goblet. In this case we pick an end-point at $y=0$ and a double root at a positive value of $y$. One then obtains a surface which we have depicted in figure \ref{fig:blackbottle}. This is then a patching of the previous black bottle solution with a disc. We leave a study of this geometry to the future since it deserves more attention than this appendix can do justice to it. 

%%%%%%%%%%%%%%%%%%%%%%%%%%%%%%%%%%%%%
\subsection{Riemann surfaces}
%%%%%%%%%%%%%%%%%%%%%%%%%%%%%%%%%%%%%

To obtain a Riemann surface we need to perform some rescalings. To this end we perform the transformations
\be
y\rightarrow \frac{\alpha}{m} +\epsilon \delta \hat{y}\, ,\quad z\rightarrow \frac{\gamma}{\epsilon}\hat{z}\, ,\quad q_i\rightarrow r_i+\frac{\alpha s\epsilon^2}{m}\, ,
\ee
with $\epsilon$ playing the role of the parameter we will eventually send to zero, and the other Greek letters constant parameters. Note that $\gamma$ and $\delta$ may be set to convenient values since they are just reparametrisations of the coordinates. 
Further that the scaling of $y$ implies that $\dd y^2\rightarrow \epsilon^2\delta^2 \dd \hat{y}^2$ and therefore to obtain a finite result we must have that $F(y)$ scales with $\epsilon^2$. This requires us to fix
\be
m^{3}r_1=\frac{2 \alpha -3\alpha^3-\alpha\sqrt{4 -9\alpha^2}}{3 }\, ,\quad m^{3}r_2=\frac{2\alpha  -3\alpha^3+\alpha\sqrt{4 -9\alpha^2}}{3}\, .\label{eq:ridef}
\ee
We may now send $\epsilon\rightarrow 0$ in the solution and obtain a finite metric and scalar fields. The scalar fields become
\be
X_1=\frac{3 \alpha^{5/4}}{2-\sqrt{4 - 9 \alpha^2}}\, ,\qquad X_2=\frac{3 \alpha^{5/4}}{2+\sqrt{4 - 9  \alpha^2}}\, , 
\ee
in particular they are constant. The metric on $M_2$ becomes
\be
\dd s^2(M_2)=\frac{ 9\alpha^2 \delta^2\dd \hat{y}^2+m^4 \gamma^2\big(4 s -3 \hat{y}^2 \delta^2(2-9  \alpha^2)\big)^2\dd\hat{z}^2}{3\alpha^2\big(4 s -3 \hat{y}^2\delta^2 (2-9  \alpha^2)\big)}\, .
\ee
Clearly for both the metric and scalar fields to be well-defined we require $4 s -3 \hat{y}^2\delta^2(2-9  \alpha^2)>0$, $4-9 \alpha^2>0$ and $2-\sqrt{4 - 9 \alpha^2}>0$. The parameter $s$ may be removed by redefinitions of the $\delta$ parameter depending on its sign and the sign of $2-9\alpha^2$. 
We now have a number of parameters to play with and the various tunings will allow us to find any of the constant curvature metrics. Note that we must take $\alpha>0$. In order to obtain a finite result for the gauge fields we must perform the constant gauge transformation,
\be
\delta A_1=\frac{3 \alpha^2}{2-\sqrt{4-9\alpha^2}}\dd z\, ,\qquad \delta A_2=\frac{3 \alpha^2}{2+\sqrt{4-9\alpha^2}}\dd z\, .
\ee

\paragraph{Two-sphere} We can find a two-sphere by taking $2-9\alpha^2>0$ which requires $s>0$. Defining
\be
\delta=\frac{2\sqrt{s}}{\sqrt{3}\sqrt{2-9\alpha^2}}\, ,\quad \gamma=\frac{\sqrt{3}\alpha}{2m^2\sqrt{s}\sqrt{2-9 \alpha^2}}\, ,\quad \hat{y}=\cos\theta\, ,
\ee
the metric becomes
\be
\dd s^2(M_2)=\frac{1}{2-9 \alpha^2}\big(\dd\theta^2+\sin^2\theta \dd\hat{z}^2\big)\, .
\ee
This gives a well-defined solution provided $0<\alpha< \frac{\sqrt{2}}{3}$. The gauge fields, after the pure gauge transformation are
\be
A_1=\frac{2-9\alpha^2+\sqrt{4-9\alpha^2}}{3 m (2-9\alpha^2)}\cos\theta\dd\hat{z}\, ,\quad A_2=\frac{2-9\alpha^2-\sqrt{4-9\alpha^2}}{3 m (2-9\alpha^2)}\cos\theta\dd\hat{z}\, .
\ee
Note that 
\be
A_{R}=g(A_1+A_2)= \cos\theta \dd\hat{z}\, ,
\ee
where we have used $2g=3m$. This is then preserving supersymmetry by the usual topological twist mechanism. To further exhibit this if one computes the Killing spinors on $M_2$ and takes this limit one finds that they become constant, see appendix \ref{app:susy} for the Killing spinors on $M_2$. The constant gauge transformations are important for this. This is of course a hallmark of a topological twist.

\paragraph{Torus} Next let us consider the two-torus case. We take $2-9\alpha^2=0$, and fix
\be 
\alpha= \frac{\sqrt{2}}{3}\, ,\quad\gamma=\frac{1}{\sqrt{6}m^2\sqrt{s}}\, ,\quad \delta=\frac{2 \sqrt{s}}{\sqrt{3}}\,,
\ee 
then the metric is 
\be
\dd s^2(M_2)=\big(\dd \hat{y}^2+\dd\hat{z}^2\big)\, ,
\ee
the scalar fields become
\be
X_1=\frac{2^{1/8}(\sqrt{2}+1)}{3^{1/4}}\, ,\quad X_2=\frac{2^{1/8}(\sqrt{2}-1)}{3^{1/4}}\, ,
\ee
and finally the gauge fields are
\be
A_1=-A_2=\frac{\sqrt{2} }{3m}\hat{y}\dd\hat{z}\, .
\ee
As in the $S^2$ case we may identify this as the topological twist since
\be
A_R=g\big(A_1+A_2\big)=0\, ,
\ee
and the Killing spinors become constant. To the best of our knowledge this solution has not appeared before in the literature before.

\paragraph{Hyperbolic space} The final case to consider is when $2-9\alpha^2<0$. Positivity of the scalar fields then fixes $\frac{\sqrt{2}}{3}<\alpha\leq \frac{2}{3}$. We have two cases depending on whether $s$ is positive or not. Firstly, let $s>0$ and then perform the redefinitions,
\be
\delta=\frac{2\sqrt{s}}{\sqrt{3}\sqrt{9 \alpha^2-2}}\, ,\quad \gamma=\frac{\sqrt{3}\alpha}{2m^2 \sqrt{s}\sqrt{9\alpha^2-02}}\, ,\quad \hat{y}=\sinh\theta\, ,
\ee
which gives the metric on $\mathbb{H}^2$ in global coordinates,
\be
\dd s^2(M_2)=\frac{1}{9\alpha^2-2}\big(\dd\theta^2+\cosh^2\theta \dd\phi^2\big)\, .
\ee
The gauge fields are
\be
A_1=\frac{2-9\alpha^2+\sqrt{4-9\alpha^2}}{3 m (9\alpha^2-2)}\cosh\theta\dd\hat{z}\, ,\quad A_2=\frac{2-9\alpha^2-\sqrt{4-9\alpha^2}}{3 m (9\alpha^2-2)}\cosh\theta\dd\hat{z}\, .
\ee
For $s<0$ we perform the redefinitions,
\be
\delta=\frac{2\sqrt{-s}}{\sqrt{3}\sqrt{9\alpha^2-2}}\, ,\quad \gamma=\frac{\sqrt{3}\alpha}{2 m^2\sqrt{-s} \sqrt{9\alpha^2-2}}\, ,\quad x=\cosh\theta\, ,
\ee
and then the metric takes the form,
\be
\dd s^2(M_2)=\frac{1}{9\alpha^2-2}\Big(\dd\theta^2 +\sinh^2\theta \dd\hat{z}^2\Big)\, ,
\ee
which is the metric on the hyperbolic plane. The resultant gauge fields are
\be
A_1=\frac{2-9\alpha^2+\sqrt{4-9\alpha^2}}{3 m (9\alpha^2-2)}\sinh\theta\dd\hat{z}\, ,\quad A_2=\frac{2-9\alpha^2-\sqrt{4-9\alpha^2}}{3 m (9\alpha^2-2)}\sinh\theta\dd\hat{z}\, .
\ee
In both cases we find
\be
A_R=g \big(A_1+A_2\big)=- A_{\mathbb{H}}\, ,\qquad \dd A_{\mathbb{H}}=\text{vol}(\mathbb{H})\, ,
\ee
and therefore we find the expected topological twist. This is the six-dimensional analogue of the Maldacena--Nunez solution for wrapped M5-branes \cite{Maldacena:2000mw}, see \cite{Nunez:2001pt} for the corresponding solutions in pure $F(4)$ gauged supergravity.

We have therefore recovered all three types of constant curvature Riemann surface from the local solution, to the best of our knowledge the $T^2$ solution is new while the others have appeared in \cite{Karndumri:2015eta}. Before we move on, it is interesting to note when it is possible for these solutions to have an enhancement of symmetry in the uplift. This is equivalent to one of the gauge fields becoming pure gauge. We see that this is not possible for the $T^2$ solution given that the two gauge fields are equal up to a sign. For the $S^2$ solution this is only possible if $\alpha=0$ at which point the metric shrinks and therefore is inadmissible. Finally, for $\mathbb{H}^2$ the condition becomes $\alpha=3^{-1/2}$, and $A_1$ vanishes. This is indeed in the permissible region for $\alpha$ and therefore this does allow for an enhancement of the symmetry. In fact it is not hard to see that this point is just taking us to the point $0=p_2(s)$ in the notation of section \ref{sec:M2sols}. A second question one may ask is if the solutions can also be found in the pure $F(4)$ theory, $i.e.$, $q_1=q_2$. From our parametrisation of the functions, this is equivalent to $r_1=r_2$ in \eqref{eq:ridef}. It is clear that there are only two solutions, either $\alpha=0$ or $\alpha=\frac{2}{3}$, the former is not admissible while the latter only makes sense for the $\mathbb{H}^2$ solution. We find the known result that only the hyperbolic horizon can be found in the pure theory \cite{Nunez:2001pt, Naka:2002jz, Kim:2019fsg}.

Note that in order to find these solutions we have taken a scaling limit. One can see what happens to the gauged metric in this limit. It is not hard to see that the fibration term in the metric $Dz^2$ becomes ungauged in this limit and the four-dimensional gauge field $\mathcal{A}$ drops out of the metric when we compactify on a Riemann surface in the truncation. One may worry that this then completely trivialises the truncation and we just end up with the AdS$_4$ vacuum solution. The answer to this is no, there are still terms in the fluxes which do not become trivial in this limit, and therefore one does end up with a consistent truncation to minimal gauged supergravity on a Riemann surface in this case too, this is the power of performing this reduction on this local solution with multiple global completions.\footnote{See \cite{Hosseini:2020wag} for a gauged version.}

%%%%%%%%%%%%%%%%%%%%%%%%%%%%%%%%%%%%%
\subsection{Domain walls/defects}
%%%%%%%%%%%%%%%%%%%%%%%%%%%%%%%%%%%%%

These solutions are the analogs of the M5-brane solutions studied in \cite{Gutperle:2022pgw} generalised to allow for more general end-points. The solutions are non-compact, but unlike the solutions with a cusp they have infinite volume, as such we cannot compute the free-energy in the same manner and obtain a sensible result. A full analysis of these solutions is beyond the scope of this paper and appendix and we leave the full analysis to the future but one should be able to mimic the analysis of \cite{Gutperle:2022pgw} with some small extensions to the allowed choice of other end-point.

%%%%%%%%%%%%%%%%%%%%%%%%%%%%%%%%%%%%%
\section{Supersymmetry of consistent truncation}\label{app:susy}
%%%%%%%%%%%%%%%%%%%%%%%%%%%%%%%%%%%%%

We now want to show that the truncation ansatz provided in the main text preserves supersymmetry. To show this, we will reduce the six-dimensional Killing spinors on the spindle down to four-dimensions and show that the resultant Killing spinor equation is the Killing spinor equation of four-dimensional minimal gauged supergravity. To begin let us lay out the conventions that we will employ in the following. The intertwiners for the SO$(1,5)$ gamma matrices satisfy
\be
A_6 \gamma_{\mu}A_6^{-1}=-\gamma_{\mu}^{\dag}\, ,\quad C_6^{-1}\gamma_{\mu}C_6=\gamma_{\mu}^{T}\, ,\quad D_6^{-1}\gamma_{\mu}D_6= -\gamma_{\mu}^{*}\, .
\ee
We will use the following gamma matrices,\footnote{Latin indices range over $\{0,1,2,3\}$ while Greek indices range over $\{ 0,1,2,3,4,5\}$.}
\begin{align}
\gamma^a=\rho^{a}\otimes \sigma^3\, ,\quad \gamma^4= 1_{4\times 4}\otimes \sigma^1\, ,\quad \gamma^5=1_{4\times 4}\otimes \sigma^2\, ,
\end{align}
where
\be
\rho^0=\ii \sigma^1\otimes \sigma^3\, ,\quad \rho^1=\sigma^2\otimes\sigma^3\, ,\quad \rho^2=1_{2\times 2}\otimes \sigma^1\, ,\quad\rho^3=1_{2\times 2}\otimes \sigma^2\,  .\label{eq:4dgamma}
\ee
The intertwiners for this choice of gamma matrices are
\be
A_6=\gamma^0\, ,\quad C_6=-\sigma^1\otimes \sigma^2\otimes \sigma^1\,,\quad D_6=C_6 A_6^T=1_{2\times 2}\otimes \sigma^1\otimes \sigma^2\, ,
\ee
and the chirality matrix is
\be
\gamma^7=\gamma^{012345}=\sigma^3\otimes\sigma^3\otimes \sigma^3\, .
\ee

\paragraph{Supersymmetry of the AdS$_4$ seed solution} Let us first set the fermionic supersymmetry transformations to vanish for the AdS$_4$ solution on which the truncation is built. The solution was shown to be supersymmetric in \cite{Faedo:2021nub}, however, we will repeat the analysis since we work with a Dirac spinor whereas they worked with the symplectic Majorana spinors. The spinor decomposes as
\be
\eta=\chi\otimes\xi\, ,\label{eq:spinortrunc}
\ee
where $\chi$ is a four-component spinor on AdS$_4$ and $\xi$ is a two component spinor on $M_2$. We take the Killing spinor on AdS$_4$ to satisfy
\be
\nabla_{m}\chi=-\frac{1}{2}\rho_{m}\chi\, ,
\ee
then we find that the gravitino equation along AdS$_4$ requires the projection condition,
\be
\bigg(m \sqrt{h_1(y) h_2(y)} 1_{8\times 8}+ \sqrt{F(y)}\gamma^{4}+\ii y^2 \gamma^{45}\bigg)\eta=0\, ,\label{eq:projection}
\ee
to be imposed. This may be reduced to a projection condition for the two-dimensional spinor on $M_2$ as
\be
\Big(m \sqrt{h_1(y)h_2(y)}\sigma^3+ \ii \sqrt{F(y)}\sigma^2 - y^2 1_{2\times 2}\Big)\xi=0\, .
\ee
We may solve the remaining conditions by fixing the spinor, $\xi$, to be\footnote{We have added in the constant gauge transformations, $\alpha_i$, of the two gauge fields in the final result. These appear in the phase of the spinor. Recall that, for the truncation, we set $\alpha_i=0$. However, when taking the Riemann surface limit, it is useful to keep this phase.}
\be
\xi=c\frac{\me^{-\ii m z\big(1-\frac{3(\alpha_1+\alpha_2)}{4}\big)}y^{1/8}}{\big(h_1(y) h_2(y)\big)^{3/16}}\begin{pmatrix}
\sqrt{F_+(y)}\\
-\sqrt{F_-(y)}
\end{pmatrix}\, ,
\ee
where
\be
F_{\pm}(y)= m \sqrt{h_1(y)h_2(y)}\pm y^{2}\, ,\qquad F_+(y)F_-(y)=F(y)\, ,
\ee
and $c$ is an arbitrary complex constant. This gives a total of eight real spinors satisfying the Killing spinor equation. Four are identified with the Poincar\'e supercharges and the remaining four are identified with the superconformal supercharges. Note that one can specialise the spinor to any of the global completions. For the Riemann surfaces, one finds constant spinors after taking the limit. This is to be expected since supersymmetry is preserved via a topological twist in these cases. 

\paragraph{Supersymmetry of the consistent truncation} In order to show that supersymmetry is preserved in the truncation we will show that the supersymmetry transformations reduce to the Killing spinor equation of four-dimensional minimal gauged supergravity,
\begin{align}
\delta\psi_{\alpha}^{4d}&=\bigg[\nabla_{\alpha}^{4d}-\frac{\ii}{2}\mathcal{A}_{\alpha}+\frac{1}{2}\rho_{\alpha}+\frac{\ii}{4}\slashed{\mathcal{F}}\cdot \gamma_{\alpha}\bigg]\chi\, ,\label{eq:4dKSE}
\end{align}
and conditions on the spinor, $\xi$, which are immediately satisfied for the seed solution.
First, let us define some notation. Let a superscript ``$g$" stand for gauging, $i.e.$,
\be
A_i^{g}=A_i\Big|_{\dd z\rightarrow Dz}\, ,
\ee
with $Dz=\dd z -\frac{1}{2m}\mathcal{A}$. Recall that the scalar fields are left unchanged in the truncation and the modification to the fluxes is only through gauging of the isometry for the gauge field and the inclusion of a two-form potential.
Let $\hat{F}_i$ be the field strength of the AdS$_4$ vacuum solution. Then we have
\be
F_i=\hat{F}_{i}^g -\frac{a_i(y)}{2m}\mathcal{F}\, ,
\ee
where
\be
a_i(y)=-\frac{y^3}{h_i(y)}\, .
\ee
Let us first consider the vanishing of $\delta \lambda$ and $\delta \chi$. The simplest is $\delta\lambda$ where we have\footnote{In the results below, all contractions will be with respect to the unwarped four-dimensional metric of four-dimensional minimal gauged supergravity. In practice this means that there are factors of $(y^2 h_1(y)h_2(y))^{1/4}$ different in $\slashed{\mathcal{F}}$. }
\be
\delta\lambda=\delta\lambda\Big|_{\mathcal{A}=0}+\frac{\ii}{4m} \Big(X_1^{-1} a_1(y)-X_2^{-1}a_{2}(y)\Big)\slashed{\mathcal{F}}\eta\, .
\ee
We have written the variation by splitting it into two pieces. The first piece, $\delta\lambda\big|_{\mathcal{A}=0}$, is equivalent to the supersymmetry variation of the AdS$_4$ seed solution if we set $\mathcal{A}=0$. It only contains the gauge field, $\mathcal{A}$, through the minimal gauging term, $Dz$, and no field strength terms. The vanishing of this term is independent of whether $\dd z$ is gauged or not and, therefore, vanishes in the truncation because it vanishes for the seed AdS$_4$ solution. The second term, on the other hand, proportional to the field strength, $\mathcal{F}$, is new and, in order to preserve supersymmetry, \emph{must} vanish identically. 
Since 
\be
a_1(y)X_2-a_2(y) X_1=0\, ,
\ee
this is satisfied. Next consider the vanishing of $\delta\chi$. We have
\begin{align}
\delta \chi=\delta \chi\Big|_{\mathcal{A}=0}+\frac{1}{16 m y^{1/4}\big(h_1(y)h_2(y)\big)^{5/8}}\slashed{\mathcal{F}}\cdot\gamma^{45}\bigg(m \sqrt{h_1(y) h_2(y)} 1_{8\times 8}+ \sqrt{F(y)}\gamma^{4}+\ii y^2 \gamma^{45}\bigg)\eta\, .
\end{align}
Using the same notation as in $\delta \lambda$, the first term vanishes identically while the second term contains the same matrix multiplying $\eta$ as in \eqref{eq:projection} and, therefore, this vanishes identically, too. Thus, the gaugini Killing spinor equations are satisfied identically. 

Finally let us consider the gravitino variation. This requires a bit more work but is still tractable to do by hand if one wishes. First we need to compute the spin connection. We will use the following frame, 
\be
e^{a}= \me^{C}\hat{e}^{a}\, ,\qquad e^4=\me^{C}\frac{y}{\sqrt{F(y)}}\dd y\, ,\qquad e^5=\me^{C}\frac{\sqrt{F(y)}}{\sqrt{h_1(y)h_2(y)}} Dz\, ,
\ee
where we used the shorthand,
\be
\me^C=\big(y^2 h_1(y)h_2(y)\big)^{1/8}\, ,
\ee
for the overall warp factor of the metric. Then we have
\begin{align}
\tensor{\omega}{^{a}_{b}}&=\tensor{\hat{\omega}}{^{a}_{b}}+\frac{F(y)}{4mh_1(y)h_2(y)}\tensor{\mathcal{F}}{^{a}_{b}}Dz\, ,\qquad &\tensor{\omega}{^{a}_{4}}&=\partial_y C \frac{\sqrt{F(y)}}{y}\hat{e}^{a}\, ,\nonumber\\
\tensor{\omega}{^{5}_{a}}&=- \frac{\sqrt{F(y)}}{4m\sqrt{h_1(y)h_2(y)}} \mathcal{F}_{ab}\hat{e}^b\, ,\qquad&\tensor{\omega}{^{4}_5}&=-\frac{\me^{-C}\sqrt{F(y)}}{y}\partial_y\bigg(\frac{\me^{C}\sqrt{F(y)}}{\sqrt{h_1(y)h_2(y)}}\bigg)Dz\, .
\end{align}

Let us now insert this into \eqref{eq:Grav6d}. First consider the $z$ components and it reduces to
\begin{align}
\delta\psi_{z}&=\delta \psi_{z}\Big|_{\mathcal{A}=0}+\frac{\sqrt{F(y)}}{16m h_1(y)h_2(y)}\slashed{\mathcal{F}}\cdot\gamma^{4}\bigg(m \sqrt{h_1(y) h_2(y)} 1_{8\times 8}+ \sqrt{F(y)}\gamma^{4}+\ii y^2 \gamma^{45}\bigg)\eta\, .
\end{align}
Once again, the bracketed part vanishes since it is the same combination that appears in \eqref{eq:projection}.
Next consider the $y$ components. Similarly we find
\be
\delta \psi_{y}=\delta \psi_y \Big|_{\mathcal{A}=0}-\frac{y}{16m \sqrt{F(y)}\sqrt{h_1(y)h_2(y)}} \slashed{\mathcal{F}}\cdot \gamma^5\bigg(m \sqrt{h_1(y) h_2(y)} 1_{8\times 8}+ \sqrt{F(y)}\gamma^{4}+\ii y^2 \gamma^{45}\bigg)\eta\, ,
\ee
and thusly the $y$ component is also immediately satisfied. 

Finally, reducing on the remaining directions, we find
\be
\delta\psi_\alpha=\delta \psi_{\alpha}^{4d}\otimes \xi\, ,
\ee
where
\be 
\delta \psi_{\alpha}^{4d}=\bigg[\nabla_{\alpha}^{4d}-\frac{\ii}{2}\mathcal{A}_{\alpha}+\frac{1}{2}\rho_{\alpha}+\frac{\ii}{4}\slashed{\mathcal{F}}\cdot \rho_{\alpha}\bigg]\chi\, .
\ee
This is the Killing spinor equation for four-dimensional minimal gauged supergravity and we, therefore, find that, given a supersymmetric solution of four-dimensional minimal gauged supergravity, we may uplift it to a supersymmetric solution of matter coupled F$(4)$ gauged supergravity on $M_2$. 

%%%%%%%%%%%%%%%%%%%%%%%%%%%%%%%%%%%%%
\section{R-symmetry of four-dimensional orbifolds}\label{app:Rsymmetry}
%%%%%%%%%%%%%%%%%%%%%%%%%%%%%%%%%%%%%

In this section, we will study the R-symmetry of the orbifolds studied in section \ref{sec:spinspin}. A similar analysis for M5-branes on a four-dimensional orbifolds was studied in \cite{Cheung:2022ilc}\footnote{A similar analysis for M2-branes on a spindle was performed in \cite{Ferrero:2020twa} }. To do this we will study the vector bilinears one can construct from the Killing spinors. As a first step we will review the vector bilinears one can construct using the Killing spinors on AdS$_2$ before studying the Killing vectors one can make with the full six-dimensional spinor, $\eta$. Recall that the Killing spinor, $\eta$, is given by the tensor product of the spinor $\chi$ on the four-dimensional spacetime and the Killing spinor on $M_2$, $\xi$, see \eqref{eq:spinortrunc}. In turn, as we will show momentarily, $\chi$ may also be decomposed into the tensor product of a spinor on AdS$_2$ and one on the spindle, $\mathbbl{\Sigma}_1$, or Riemann surface. Here we will focus only on the spindle solution since this has novel features. 

The four-dimensional Killing spinor satisfying \eqref{eq:4dKSE} is
\be
\chi=\theta^{\text{AdS}_2}\otimes \epsilon\,,
\ee
where
\be
\epsilon=\frac{1}{\sqrt{x}}\begin{pmatrix}
\sqrt{q_-(x)}\\
-\sqrt{q_{+}(x)}\end{pmatrix}\, .
\ee
We have defined
\be
q_{\pm}(x)=x^2 \pm (2x-\mathtt{a})\, ,\qquad q(x)=q_+(x)q_-(x)\, ,
\ee
with $q(x)$ defined in \eqref{eq:q(x)def} and $\theta^{\text{AdS}}$ satisfies 
\be
\nabla^{\text{AdS}_2}_{\alpha}\theta^{\text{AdS}_2}=\frac{1}{2}\Gamma_{\alpha}\theta^{\text{AdS}_2}\, .\label{eq:AdS2KSE}
\ee

First we must give the Killing spinors of AdS$_2$. We take the gamma matrices for AdS$_2$ to be
\be
\Gamma^0=\ii \sigma^1\, ,\qquad \Gamma^1=\sigma^2\, ,
\ee
which is compatible with the gamma matrices we chose in the previous section, \eqref{eq:4dgamma}. We take the metric on AdS$_2$ to be
\be
\dd s^2=r^2\dd t^2+\frac{\dd r^2}{r^2}\, .
\ee
Then the Killing spinors on AdS$_2$ satisfying \eqref{eq:AdS2KSE}
are
\be
\theta_1=\frac{\sqrt{r}}{2}\begin{pmatrix} 1\\ \ii\end{pmatrix}\, ,\quad \theta_2=\frac{1}{\sqrt{2}\sqrt{r}}\big(1_{2\times 2}-t\,  r\,  \sigma^3\big)\begin{pmatrix}1\\-\ii \end{pmatrix}\, .
\ee
We may construct three Killing vectors from these Killing spinors which generate the SO$(1,1)$ isometry of the spacetime. We have
\begin{align}
k_{11}=\bar{\theta}_1\Gamma^{\alpha}\theta_1\partial^{\alpha}&=-\partial_t\, ,\nonumber\\
k_{22}=\bar{\theta}_2\Gamma^{\alpha}\theta_2\partial_{\alpha}&=-\frac{1+r^2 t^2}{r^2}\partial_t+2 t r \partial_r\, ,\\
k_{12}=\bar{\theta}_1\Gamma^{\alpha}\theta_2\partial_{\alpha}&=t \partial_t -r \partial_r\nonumber\, .
\end{align}
These satisfy the expected commutation relations,
\be
[k_{11},k_{22}]=2 k_{12}\, ,\quad [k_{11},k_{12}]=k_{11}\, ,\quad [k_{22},k_{12}]=-k_{22}\, .
\ee

We can now construct the analogous six-dimensional vector bilinears. Let us define
\be
\eta_1=\theta_1\otimes \epsilon\otimes \xi\, ,\quad\eta_2=\theta_2\otimes \epsilon\otimes \xi\, ,
\ee
and we have
\begin{align}
K_{11}=&\bar{\eta_1}\gamma^{\mu}\eta_1\partial_{\mu}=m k_{11}\, ,\nonumber\\
K_{22}=&\bar{\eta_2}\gamma^{\mu}\eta_2\partial_{\mu}=m k_{22}\, ,\\
K_{12}=&\bar{\eta_1}\gamma^{\mu}\eta_2\partial_{\mu}= m k_{12}+\ii \Big(m\partial_{\phi}-\frac{1}{2}\partial_{z}\Big)\, .\nonumber
\end{align}
This generates the superconformal algebra of 1d $\mathcal{N}=2$ superconformal quantum mechanical theory which is dual to 5d USp$(2N)$ theory living on a stack of D4-D8-branes wrapped on the four-dimensional orbifold. Note that the R-symmetry, appearing as the complex part of the $K_{12}$ bilinear, is a linear combination of the two U$(1)$'s,
\be
R=m \partial_{\phi}-\frac{1}{2}\partial_{z}\, .
\ee

%%%%%%%%%%%%%%%%%%%%%%%%%%%%%%%%%%%%%

\vspace{5cm}

\bibliographystyle{JHEP}
\bibliography{reference}

\providecommand{\href}[2]{#2}\begingroup\raggedright\begin{thebibliography}{10}

\bibitem{Witten:1988ze}
E.~Witten, \emph{{Topological Quantum Field Theory}},
  \href{http://dx.doi.org/10.1007/BF01223371}{\emph{Commun. Math. Phys.} {\bf
  117} (1988) 353}.

\bibitem{Maldacena:2000mw}
J.~M. Maldacena and C.~Nunez, \emph{{Supergravity description of field theories
  on curved manifolds and a no go theorem}},
  \href{http://dx.doi.org/10.1142/S0217751X01003937}{\emph{Int. J. Mod. Phys.
  A} {\bf 16} (2001) 822--855},
  [\href{https://arxiv.org/abs/hep-th/0007018}{{\tt hep-th/0007018}}].

\bibitem{Ferrero:2020laf}
P.~Ferrero, J.~P. Gauntlett, J.~M. P\'erez Ipi\~na, D.~Martelli and J.~Sparks,
  \emph{{D3-Branes Wrapped on a Spindle}},
  \href{http://dx.doi.org/10.1103/PhysRevLett.126.111601}{\emph{Phys. Rev.
  Lett.} {\bf 126} (2021) 111601},
  [\href{https://arxiv.org/abs/2011.10579}{{\tt 2011.10579}}].

\bibitem{Hosseini:2021fge}
S.~M. Hosseini, K.~Hristov and A.~Zaffaroni, \emph{{Rotating multi-charge
  spindles and their microstates}},
  \href{http://dx.doi.org/10.1007/JHEP07(2021)182}{\emph{JHEP} {\bf 07} (2021)
  182}, [\href{https://arxiv.org/abs/2104.11249}{{\tt 2104.11249}}].

\bibitem{Boido:2021szx}
A.~Boido, J.~M.~P. Ipi\~na and J.~Sparks, \emph{{Twisted D3-brane and M5-brane
  compactifications from multi-charge spindles}},
  \href{http://dx.doi.org/10.1007/JHEP07(2021)222}{\emph{JHEP} {\bf 07} (2021)
  222}, [\href{https://arxiv.org/abs/2104.13287}{{\tt 2104.13287}}].

\bibitem{Ferrero:2020twa}
P.~Ferrero, J.~P. Gauntlett, J.~M.~P. Ipi\~na, D.~Martelli and J.~Sparks,
  \emph{{Accelerating black holes and spinning spindles}},
  \href{http://dx.doi.org/10.1103/PhysRevD.104.046007}{\emph{Phys. Rev. D} {\bf
  104} (2021) 046007}, [\href{https://arxiv.org/abs/2012.08530}{{\tt
  2012.08530}}].

\bibitem{Cassani:2021dwa}
D.~Cassani, J.~P. Gauntlett, D.~Martelli and J.~Sparks, \emph{{Thermodynamics
  of accelerating and supersymmetric AdS4 black holes}},
  \href{http://dx.doi.org/10.1103/PhysRevD.104.086005}{\emph{Phys. Rev. D} {\bf
  104} (2021) 086005}, [\href{https://arxiv.org/abs/2106.05571}{{\tt
  2106.05571}}].

\bibitem{Ferrero:2021ovq}
P.~Ferrero, M.~Inglese, D.~Martelli and J.~Sparks, \emph{{Multicharge
  accelerating black holes and spinning spindles}},
  \href{http://dx.doi.org/10.1103/PhysRevD.105.126001}{\emph{Phys. Rev. D} {\bf
  105} (2022) 126001}, [\href{https://arxiv.org/abs/2109.14625}{{\tt
  2109.14625}}].

\bibitem{Couzens:2021rlk}
C.~Couzens, K.~Stemerdink and D.~van~de Heisteeg, \emph{{M2-branes on discs and
  multi-charged spindles}},
  \href{http://dx.doi.org/10.1007/JHEP04(2022)107}{\emph{JHEP} {\bf 04} (2022)
  107}, [\href{https://arxiv.org/abs/2110.00571}{{\tt 2110.00571}}].

\bibitem{Faedo:2021kur}
F.~Faedo, S.~Klemm and A.~Vigano, \emph{{Supersymmetric black holes with spiky
  horizons}}, \href{http://dx.doi.org/10.1007/JHEP09(2021)102}{\emph{JHEP} {\bf
  09} (2021) 102}, [\href{https://arxiv.org/abs/2105.02902}{{\tt 2105.02902}}].

\bibitem{Ferrero:2021wvk}
P.~Ferrero, J.~P. Gauntlett, D.~Martelli and J.~Sparks, \emph{{M5-branes
  wrapped on a spindle}},
  \href{http://dx.doi.org/10.1007/JHEP11(2021)002}{\emph{JHEP} {\bf 11} (2021)
  002}, [\href{https://arxiv.org/abs/2105.13344}{{\tt 2105.13344}}].

\bibitem{Faedo:2021nub}
F.~Faedo and D.~Martelli, \emph{{D4-branes wrapped on a spindle}},
  \href{http://dx.doi.org/10.1007/JHEP02(2022)101}{\emph{JHEP} {\bf 02} (2022)
  101}, [\href{https://arxiv.org/abs/2111.13660}{{\tt 2111.13660}}].

\bibitem{Giri:2021xta}
S.~Giri, \emph{{Black holes with spindles at the horizon}},
  \href{http://dx.doi.org/10.1007/JHEP06(2022)145}{\emph{JHEP} {\bf 06} (2022)
  145}, [\href{https://arxiv.org/abs/2112.04431}{{\tt 2112.04431}}].

\bibitem{Arav:2022lzo}
I.~Arav, J.~P. Gauntlett, M.~M. Roberts and C.~Rosen, \emph{{Leigh-Strassler
  compactified on a spindle}},
  \href{http://dx.doi.org/10.1007/JHEP10(2022)067}{\emph{JHEP} {\bf 10} (2022)
  067}, [\href{https://arxiv.org/abs/2207.06427}{{\tt 2207.06427}}].

\bibitem{Couzens:2022yiv}
C.~Couzens and K.~Stemerdink, \emph{{Universal spindles: D2's on $\Sigma$ and
  M5's on $\Sigma\times \mathbb{H}^3$}},
  \href{https://arxiv.org/abs/2207.06449}{{\tt 2207.06449}}.

\bibitem{Cheung:2022wpg}
K.~C.~M. Cheung and R.~Leung, \emph{{Type IIA embeddings of D = 5 minimal
  gauged supergravity via non-Abelian T-duality}},
  \href{http://dx.doi.org/10.1007/JHEP06(2022)051}{\emph{JHEP} {\bf 06} (2022)
  051}, [\href{https://arxiv.org/abs/2203.15114}{{\tt 2203.15114}}].

\bibitem{Couzens:2022aki}
C.~Couzens, N.~T. Macpherson and A.~Passias, \emph{{A plethora of Type IIA
  embeddings for $d=5$ minimal supergravity}},
  \href{https://arxiv.org/abs/2209.15540}{{\tt 2209.15540}}.

\bibitem{Ferrero:2021etw}
P.~Ferrero, J.~P. Gauntlett and J.~Sparks, \emph{{Supersymmetric spindles}},
  \href{http://dx.doi.org/10.1007/JHEP01(2022)102}{\emph{JHEP} {\bf 01} (2022)
  102}, [\href{https://arxiv.org/abs/2112.01543}{{\tt 2112.01543}}].

\bibitem{Couzens:2021cpk}
C.~Couzens, \emph{{A tale of (M)2 twists}},
  \href{http://dx.doi.org/10.1007/JHEP03(2022)078}{\emph{JHEP} {\bf 03} (2022)
  078}, [\href{https://arxiv.org/abs/2112.04462}{{\tt 2112.04462}}].

\bibitem{Bah:2021mzw}
I.~Bah, F.~Bonetti, R.~Minasian and E.~Nardoni, \emph{{Holographic Duals of
  Argyres-Douglas Theories}},
  \href{http://dx.doi.org/10.1103/PhysRevLett.127.211601}{\emph{Phys. Rev.
  Lett.} {\bf 127} (2021) 211601},
  [\href{https://arxiv.org/abs/2105.11567}{{\tt 2105.11567}}].

\bibitem{Bah:2021hei}
I.~Bah, F.~Bonetti, R.~Minasian and E.~Nardoni, \emph{{M5-brane sources,
  holography, and Argyres-Douglas theories}},
  \href{http://dx.doi.org/10.1007/JHEP11(2021)140}{\emph{JHEP} {\bf 11} (2021)
  140}, [\href{https://arxiv.org/abs/2106.01322}{{\tt 2106.01322}}].

\bibitem{Argyres:1995jj}
P.~C. Argyres and M.~R. Douglas, \emph{{New phenomena in SU(3) supersymmetric
  gauge theory}},
  \href{http://dx.doi.org/10.1016/0550-3213(95)00281-V}{\emph{Nucl. Phys. B}
  {\bf 448} (1995) 93--126}, [\href{https://arxiv.org/abs/hep-th/9505062}{{\tt
  hep-th/9505062}}].

\bibitem{Couzens:2022yjl}
C.~Couzens, H.~Kim, N.~Kim and Y.~Lee, \emph{{Holographic duals of M5-branes on
  an irregularly punctured sphere}},
  \href{http://dx.doi.org/10.1007/JHEP07(2022)102}{\emph{JHEP} {\bf 07} (2022)
  102}, [\href{https://arxiv.org/abs/2204.13537}{{\tt 2204.13537}}].

\bibitem{Bah:2022yjf}
I.~Bah, F.~Bonetti, E.~Nardoni and T.~Waddleton, \emph{{Aspects of Irregular
  Punctures via Holography}},  \href{https://arxiv.org/abs/2207.10094}{{\tt
  2207.10094}}.

\bibitem{Couzens:2021tnv}
C.~Couzens, N.~T. Macpherson and A.~Passias, \emph{{$ \mathcal{N} $ = (2, 2)
  AdS$_{3}$ from D3-branes wrapped on Riemann surfaces}},
  \href{http://dx.doi.org/10.1007/JHEP02(2022)189}{\emph{JHEP} {\bf 02} (2022)
  189}, [\href{https://arxiv.org/abs/2107.13562}{{\tt 2107.13562}}].

\bibitem{Suh:2021ifj}
M.~Suh, \emph{{D3-branes and M5-branes wrapped on a topological disc}},
  \href{http://dx.doi.org/10.1007/JHEP03(2022)043}{\emph{JHEP} {\bf 03} (2022)
  043}, [\href{https://arxiv.org/abs/2108.01105}{{\tt 2108.01105}}].

\bibitem{Suh:2021hef}
M.~Suh, \emph{{M2-branes wrapped on a topological disk}},
  \href{http://dx.doi.org/10.1007/JHEP09(2022)048}{\emph{JHEP} {\bf 09} (2022)
  048}, [\href{https://arxiv.org/abs/2109.13278}{{\tt 2109.13278}}].

\bibitem{Suh:2021aik}
M.~Suh, \emph{{D4-branes wrapped on a topological disk}},
  \href{https://arxiv.org/abs/2108.08326}{{\tt 2108.08326}}.

\bibitem{Karndumri:2022wpu}
P.~Karndumri and P.~Nuchino, \emph{{Five-branes wrapped on topological disks
  from 7D N=2 gauged supergravity}},
  \href{http://dx.doi.org/10.1103/PhysRevD.105.066010}{\emph{Phys. Rev. D} {\bf
  105} (2022) 066010}, [\href{https://arxiv.org/abs/2201.05037}{{\tt
  2201.05037}}].

\bibitem{Gutperle:2022pgw}
M.~Gutperle and N.~Klein, \emph{{A note on co-dimension 2 defects in N=4,d=7
  gauged supergravity}},
  \href{http://dx.doi.org/10.1016/j.nuclphysb.2022.115969}{\emph{Nucl. Phys. B}
  {\bf 984} (2022) 115969}, [\href{https://arxiv.org/abs/2203.13839}{{\tt
  2203.13839}}].

\bibitem{Suh:2022olh}
M.~Suh, \emph{{M5-branes and D4-branes wrapped on a direct product of spindle
  and Riemann surface}},  \href{https://arxiv.org/abs/2207.00034}{{\tt
  2207.00034}}.

\bibitem{Suh:2018tul}
M.~Suh, \emph{{Supersymmetric AdS$_{6}$ black holes from F(4) gauged
  supergravity}}, \href{http://dx.doi.org/10.1007/JHEP01(2019)035}{\emph{JHEP}
  {\bf 01} (2019) 035}, [\href{https://arxiv.org/abs/1809.03517}{{\tt
  1809.03517}}].

\bibitem{Hosseini:2018usu}
S.~M. Hosseini, K.~Hristov, A.~Passias and A.~Zaffaroni, \emph{{6D attractors
  and black hole microstates}},
  \href{http://dx.doi.org/10.1007/JHEP12(2018)001}{\emph{JHEP} {\bf 12} (2018)
  001}, [\href{https://arxiv.org/abs/1809.10685}{{\tt 1809.10685}}].

\bibitem{Suh:2018szn}
M.~Suh, \emph{{Supersymmetric $AdS_6$ black holes from matter coupled $F(4)$
  gauged supergravity}},
  \href{http://dx.doi.org/10.1007/JHEP02(2019)108}{\emph{JHEP} {\bf 02} (2019)
  108}, [\href{https://arxiv.org/abs/1810.00675}{{\tt 1810.00675}}].

\bibitem{Gauntlett:2001jj}
J.~P. Gauntlett and N.~Kim, \emph{{M five-branes wrapped on supersymmetric
  cycles. 2.}}, \href{http://dx.doi.org/10.1103/PhysRevD.65.086003}{\emph{Phys.
  Rev. D} {\bf 65} (2002) 086003},
  [\href{https://arxiv.org/abs/hep-th/0109039}{{\tt hep-th/0109039}}].

\bibitem{Cheung:2022ilc}
K.~C.~M. Cheung, J.~H.~T. Fry, J.~P. Gauntlett and J.~Sparks, \emph{{M5-branes
  wrapped on four-dimensional orbifolds}},
  \href{http://dx.doi.org/10.1007/JHEP08(2022)082}{\emph{JHEP} {\bf 08} (2022)
  082}, [\href{https://arxiv.org/abs/2204.02990}{{\tt 2204.02990}}].

\bibitem{Brandhuber:1999np}
A.~Brandhuber and Y.~Oz, \emph{{The D-4 - D-8 brane system and five-dimensional
  fixed points}},
  \href{http://dx.doi.org/10.1016/S0370-2693(99)00763-7}{\emph{Phys. Lett. B}
  {\bf 460} (1999) 307--312}, [\href{https://arxiv.org/abs/hep-th/9905148}{{\tt
  hep-th/9905148}}].

\bibitem{Cvetic:1999xx}
M.~Cvetic, S.~S. Gubser, H.~Lu and C.~N. Pope, \emph{{Symmetric potentials of
  gauged supergravities in diverse dimensions and Coulomb branch of gauge
  theories}}, \href{http://dx.doi.org/10.1103/PhysRevD.62.086003}{\emph{Phys.
  Rev. D} {\bf 62} (2000) 086003},
  [\href{https://arxiv.org/abs/hep-th/9909121}{{\tt hep-th/9909121}}].

\bibitem{Faedo:2022aaa}
F.~Faedo, A.~Fontanarossa and D.~Martelli, \emph{{Branes wrapped on orbifolds
  and their gravitational blocks}}, .

\bibitem{Romans:1985tw}
L.~J. Romans, \emph{{The F(4) Gauged Supergravity in Six-dimensions}},
  \href{http://dx.doi.org/10.1016/0550-3213(86)90517-1}{\emph{Nucl. Phys. B}
  {\bf 269} (1986) 691}.

\bibitem{Andrianopoli:2001rs}
L.~Andrianopoli, R.~D'Auria and S.~Vaula, \emph{{Matter coupled F(4) gauged
  supergravity Lagrangian}},
  \href{http://dx.doi.org/10.1088/1126-6708/2001/05/065}{\emph{JHEP} {\bf 05}
  (2001) 065}, [\href{https://arxiv.org/abs/hep-th/0104155}{{\tt
  hep-th/0104155}}].

\bibitem{Karndumri:2015eta}
P.~Karndumri, \emph{{Twisted compactification of N = 2 5D SCFTs to three and
  two dimensions from F(4) gauged supergravity}},
  \href{http://dx.doi.org/10.1007/JHEP09(2015)034}{\emph{JHEP} {\bf 09} (2015)
  034}, [\href{https://arxiv.org/abs/1507.01515}{{\tt 1507.01515}}].

\bibitem{Suh:2020rma}
M.~Suh, \emph{{The non-SUSY AdS$_{6}$ and AdS$_{7}$ fixed points are brane-jet
  unstable}}, \href{http://dx.doi.org/10.1007/JHEP10(2020)010}{\emph{JHEP} {\bf
  10} (2020) 010}, [\href{https://arxiv.org/abs/2004.06823}{{\tt 2004.06823}}].

\bibitem{Cvetic:1999un}
M.~Cvetic, H.~Lu and C.~N. Pope, \emph{{Gauged six-dimensional supergravity
  from massive type IIA}},
  \href{http://dx.doi.org/10.1103/PhysRevLett.83.5226}{\emph{Phys. Rev. Lett.}
  {\bf 83} (1999) 5226--5229},
  [\href{https://arxiv.org/abs/hep-th/9906221}{{\tt hep-th/9906221}}].

\bibitem{Seiberg:1996bd}
N.~Seiberg, \emph{{Five-dimensional SUSY field theories, nontrivial fixed
  points and string dynamics}},
  \href{http://dx.doi.org/10.1016/S0370-2693(96)01215-4}{\emph{Phys. Lett. B}
  {\bf 388} (1996) 753--760}, [\href{https://arxiv.org/abs/hep-th/9608111}{{\tt
  hep-th/9608111}}].

\bibitem{Malek:2019ucd}
E.~Malek, H.~Samtleben and V.~Vall~Camell, \emph{{Supersymmetric AdS$_7$ and
  AdS$_6$ vacua and their consistent truncations with vector multiplets}},
  \href{http://dx.doi.org/10.1007/JHEP04(2019)088}{\emph{JHEP} {\bf 04} (2019)
  088}, [\href{https://arxiv.org/abs/1901.11039}{{\tt 1901.11039}}].

\bibitem{Jeong:2013jfc}
J.~Jeong, O.~Kelekci and E.~O~Colgain, \emph{{An alternative IIB embedding of
  F(4) gauged supergravity}},
  \href{http://dx.doi.org/10.1007/JHEP05(2013)079}{\emph{JHEP} {\bf 05} (2013)
  079}, [\href{https://arxiv.org/abs/1302.2105}{{\tt 1302.2105}}].

\bibitem{Hong:2018amk}
J.~Hong, J.~T. Liu and D.~R. Mayerson, \emph{{Gauged Six-Dimensional
  Supergravity from Warped IIB Reductions}},
  \href{http://dx.doi.org/10.1007/JHEP09(2018)140}{\emph{JHEP} {\bf 09} (2018)
  140}, [\href{https://arxiv.org/abs/1808.04301}{{\tt 1808.04301}}].

\bibitem{Malek:2018zcz}
E.~Malek, H.~Samtleben and V.~Vall~Camell, \emph{{Supersymmetric AdS$_{7}$ and
  AdS$_6$ vacua and their minimal consistent truncations from exceptional field
  theory}}, \href{http://dx.doi.org/10.1016/j.physletb.2018.09.037}{\emph{Phys.
  Lett. B} {\bf 786} (2018) 171--179},
  [\href{https://arxiv.org/abs/1808.05597}{{\tt 1808.05597}}].

\bibitem{Faedo:2019cvr}
A.~F. Faedo, C.~Nunez and C.~Rosen, \emph{{Consistent truncations of
  supergravity and $\frac{1}{2}$-BPS RG flows in $4d$ SCFTs}},
  \href{http://dx.doi.org/10.1007/JHEP03(2020)080}{\emph{JHEP} {\bf 03} (2020)
  080}, [\href{https://arxiv.org/abs/1912.13516}{{\tt 1912.13516}}].

\bibitem{Hosseini:2020wag}
S.~M. Hosseini and K.~Hristov, \emph{{4d F(4) gauged supergravity and black
  holes of class $\mathcal{F}$}},
  \href{http://dx.doi.org/10.1007/JHEP02(2021)177}{\emph{JHEP} {\bf 02} (2021)
  177}, [\href{https://arxiv.org/abs/2011.01943}{{\tt 2011.01943}}].

\bibitem{Romans:1991nq}
L.~J. Romans, \emph{{Supersymmetric, cold and lukewarm black holes in
  cosmological Einstein-Maxwell theory}},
  \href{http://dx.doi.org/10.1016/0550-3213(92)90684-4}{\emph{Nucl. Phys. B}
  {\bf 383} (1992) 395--415}, [\href{https://arxiv.org/abs/hep-th/9203018}{{\tt
  hep-th/9203018}}].

\bibitem{Caldarelli:1998hg}
M.~M. Caldarelli and D.~Klemm, \emph{{Supersymmetry of Anti-de Sitter black
  holes}}, \href{http://dx.doi.org/10.1016/S0550-3213(98)00846-3}{\emph{Nucl.
  Phys. B} {\bf 545} (1999) 434--460},
  [\href{https://arxiv.org/abs/hep-th/9808097}{{\tt hep-th/9808097}}].

\bibitem{Maldacena:1997re}
J.~M. Maldacena, \emph{{The Large N limit of superconformal field theories and
  supergravity}}, \href{http://dx.doi.org/10.1023/A:1026654312961}{\emph{Adv.
  Theor. Math. Phys.} {\bf 2} (1998) 231--252},
  [\href{https://arxiv.org/abs/hep-th/9711200}{{\tt hep-th/9711200}}].

\bibitem{Hosseini:2018uzp}
S.~M. Hosseini, I.~Yaakov and A.~Zaffaroni, \emph{{Topologically twisted
  indices in five dimensions and holography}},
  \href{http://dx.doi.org/10.1007/JHEP11(2018)119}{\emph{JHEP} {\bf 11} (2018)
  119}, [\href{https://arxiv.org/abs/1808.06626}{{\tt 1808.06626}}].

\bibitem{Crichigno:2018adf}
P.~M. Crichigno, D.~Jain and B.~Willett, \emph{{5d Partition Functions with A
  Twist}}, \href{http://dx.doi.org/10.1007/JHEP11(2018)058}{\emph{JHEP} {\bf
  11} (2018) 058}, [\href{https://arxiv.org/abs/1808.06744}{{\tt 1808.06744}}].

\bibitem{Suh:2019ily}
M.~Suh, \emph{{Uplifting supersymmetric AdS$_{6}$ black holes to type II
  supergravity}}, \href{http://dx.doi.org/10.1007/JHEP12(2020)059}{\emph{JHEP}
  {\bf 12} (2020) 059}, [\href{https://arxiv.org/abs/1908.09846}{{\tt
  1908.09846}}].

\bibitem{Romans:1985tz}
L.~J. Romans, \emph{{Massive N=2a Supergravity in Ten-Dimensions}},
  \href{http://dx.doi.org/10.1016/0370-2693(86)90375-8}{\emph{Phys. Lett. B}
  {\bf 169} (1986) 374}.

\bibitem{Bah:2017wxp}
I.~Bah, A.~Passias and A.~Tomasiello, \emph{{AdS$_{5}$ compactifications with
  punctures in massive IIA supergravity}},
  \href{http://dx.doi.org/10.1007/JHEP11(2017)050}{\emph{JHEP} {\bf 11} (2017)
  050}, [\href{https://arxiv.org/abs/1704.07389}{{\tt 1704.07389}}].

\bibitem{Passias:2018zlm}
A.~Passias, D.~Prins and A.~Tomasiello, \emph{{A massive class of $\mathcal{N}
  = 2$ AdS$_4$ IIA solutions}},
  \href{http://dx.doi.org/10.1007/JHEP10(2018)071}{\emph{JHEP} {\bf 10} (2018)
  071}, [\href{https://arxiv.org/abs/1805.03661}{{\tt 1805.03661}}].

\bibitem{Klemm:2014rda}
D.~Klemm, \emph{{Four-dimensional black holes with unusual horizons}},
  \href{http://dx.doi.org/10.1103/PhysRevD.89.084007}{\emph{Phys. Rev. D} {\bf
  89} (2014) 084007}, [\href{https://arxiv.org/abs/1401.3107}{{\tt
  1401.3107}}].

\bibitem{Hennigar:2015cja}
R.~A. Hennigar, D.~Kubiz\v{n}\'ak, R.~B. Mann and N.~Musoke,
  \emph{{Ultraspinning limits and super-entropic black holes}},
  \href{http://dx.doi.org/10.1007/JHEP06(2015)096}{\emph{JHEP} {\bf 06} (2015)
  096}, [\href{https://arxiv.org/abs/1504.07529}{{\tt 1504.07529}}].

\bibitem{Gnecchi:2013mja}
A.~Gnecchi, K.~Hristov, D.~Klemm, C.~Toldo and O.~Vaughan, \emph{{Rotating
  black holes in 4d gauged supergravity}},
  \href{http://dx.doi.org/10.1007/JHEP01(2014)127}{\emph{JHEP} {\bf 01} (2014)
  127}, [\href{https://arxiv.org/abs/1311.1795}{{\tt 1311.1795}}].

\bibitem{Chen:2016rjt}
Y.~Chen and E.~Teo, \emph{{Black holes with bottle-shaped horizons}},
  \href{http://dx.doi.org/10.1103/PhysRevD.93.124028}{\emph{Phys. Rev. D} {\bf
  93} (2016) 124028}, [\href{https://arxiv.org/abs/1604.07527}{{\tt
  1604.07527}}].

\bibitem{Chen:2015zoa}
Y.~Chen, Y.-K. Lim and E.~Teo, \emph{{Deformed hyperbolic black holes}},
  \href{http://dx.doi.org/10.1103/PhysRevD.92.044058}{\emph{Phys. Rev. D} {\bf
  92} (2015) 044058}, [\href{https://arxiv.org/abs/1507.02416}{{\tt
  1507.02416}}].

\bibitem{Nunez:2001pt}
C.~Nunez, I.~Y. Park, M.~Schvellinger and T.~A. Tran, \emph{{Supergravity duals
  of gauge theories from F(4) gauged supergravity in six-dimensions}},
  \href{http://dx.doi.org/10.1088/1126-6708/2001/04/025}{\emph{JHEP} {\bf 04}
  (2001) 025}, [\href{https://arxiv.org/abs/hep-th/0103080}{{\tt
  hep-th/0103080}}].

\bibitem{Naka:2002jz}
M.~Naka, \emph{{Various wrapped branes from gauged supergravities}},
  \href{https://arxiv.org/abs/hep-th/0206141}{{\tt hep-th/0206141}}.

\bibitem{Kim:2019fsg}
N.~Kim and M.~Shim, \emph{{Wrapped Brane Solutions in Romans $F(4)$ Gauged
  Supergravity}},
  \href{http://dx.doi.org/10.1016/j.nuclphysb.2019.114882}{\emph{Nucl. Phys. B}
  {\bf 951} (2020) 114882}, [\href{https://arxiv.org/abs/1909.01534}{{\tt
  1909.01534}}].

\end{thebibliography}\endgroup

%%%%%%%%%%%%%%%%%%%%%%%%%%%%%%%%%%%%%%%%%%%%%%%%%%%%%%%%%%%%%
\end{document}